\documentclass[aps,prb,showpacs,reprint,superscriptaddress,lengthcheck]{revtex4-1}
\usepackage{xcolor,colortbl}
\usepackage[english]{babel}
\usepackage{amsmath}
\usepackage{amssymb, comment}
\usepackage[]{graphicx}
\usepackage{times}
\usepackage{bm}
\usepackage{units}
\usepackage{braket}
\usepackage[version=3]{mhchem}

\usepackage{bbm}
\usepackage{tabularx}

\usepackage{color}

\renewcommand{\eqref}[1]{Eq. (\ref{#1})}

\begin{document}
\title{Interlayer exciton landscape in WS$_2$/tetracene heterostructures }

\author{Joshua J. P. Thompson} 
\affiliation{Department of Physics, Philipps-Universit\"{a}t Marburg,
35037 Marburg, Germany}
\email{joshua.thompson@physik.uni-marburg.de}
\author{Victoria Lumsargis}
\affiliation{Department of Chemistry, Purdue University, West Lafayette, Indiana, 47907, United States}
\author{Maja Feierabend}
\affiliation{Department of 
Physics, Chalmers University of Technology,  412 96 Gothenburg, Sweden}
\author{Quichen Zhao}
\affiliation{Department of Chemistry, Purdue University, West Lafayette, Indiana, 47907, United States}
\affiliation{State Key Laboratory of Superhard Materials, Jilin University, Changchun, Jilin, 130012, China}
\author{Kang Wang}
\affiliation{Davidson School of Chemical Engineering,  Purdue University, West Lafayette, Indiana,  47907, United States}
\author{Letian Dou}
\affiliation{Davidson School of Chemical Engineering,  Purdue University, West Lafayette, Indiana,  47907, United States}
\author{Libai Huang}
\affiliation{Department of Chemistry, Purdue University, West Lafayette, Indiana, 47907, United States}
\author{Ermin Malic}
\affiliation{Department of Physics, Philipps-Universit\"{a}t Marburg,
35037 Marburg, Germany}
\affiliation{Department of 
Physics, Chalmers University of Technology,  412 96 Gothenburg, Sweden}


\begin{abstract}
The vertical stacking of two-dimensional materials into heterostructures gives rise to a plethora of intriguing optoelectronic properties and presents an unprecedented potential for technological development. While much progress has been made combining different monolayers of transition metal dichalgonenides (TMDs), little is known about TMD-based heterostructures including organic layers of molecules. 
Here, we present a joint theory-experiment study on a TMD/tetracene heterostructure demonstrating clear signatures of spatially separated interlayer excitons in low temperature photoluminescence spectra. Here, the Coulomb-bound electrons and holes are localized either in the TMD or in the molecule layer, respectively.  In particular, we reveal both in theory and experiment signatures of the entire intra- and interlayer exciton landscape in the photoluminescence spectra. In particular, we find both in theory and experiment a pronounced transfer of intensity from the intralayer TMD exciton to a series of energetically lower interlayer excitons with decreasing temperature. In addition, we find signatures phonon-sidebands stemming from these interlayer exciton states. Our findings shed light on the microscopic nature of interlayer excitons in TMD/molecule heterostructures and could have important implications for technological applications of these materials. 
\end{abstract}

\maketitle
%
%
A controlled vertical stacking of atomically thin materials into van der Waals heterostructures with tailored properties has recently become feasible \cite{geim2013van, liu2019recent}. In particular, there has been much progress in understanding homo- and heterobilayers of transition metal dichalcogenides (TMDs) \cite{alexeev2019resonantly, jin2019observation, Tran2019EvidenceFM, merkl2019ultrafast, brem2020hybridized, brem2020tunable}.
The strong Coulomb interaction leads to a variety of excitons in these materials  \cite{deilmann2019finite, malic2018dark,madeo2020directly,wallauer2021momentum}. Recent experiments have observed signatures of spatially separated interlayer exciton states (ILXs) \cite{rivera2015observation,merkl2019ultrafast,hong2014ultrafast, kunstmann2018momentum}. So far, ILX signatures have been mostly demonstrated in TMD homo- and heterobilayers. A recent study showed that ILXs can also be observed in organic/heterostructures, where a thin crystalline film of organic molecules is stacked on a TMD monolayer \cite{zhu2018highly}. Tetracene (Tc) molecules are of particular interest due to their excellent light-emitting and light-harvesting properties  \cite{li2020recent,gobbi20182d,lee2014heterostructures} and  their similar band gap energies to TMD materials \cite{costa2016optical}.  

\begin{figure}[t!]
\includegraphics[width=\linewidth]{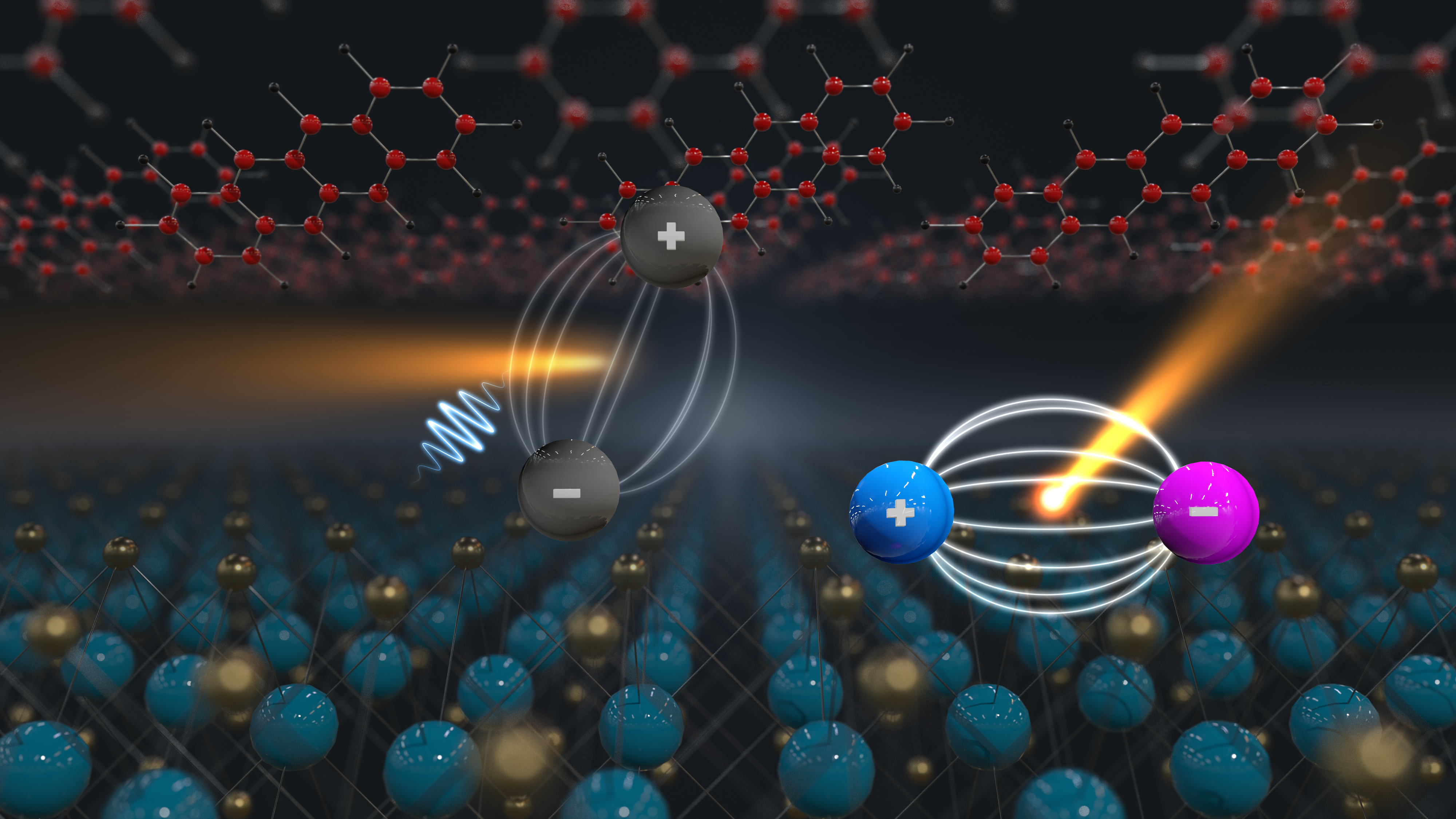} 
\caption{\textbf{Artistic illustration of intra- and interlayer excitons in a TMD/Tc heterostructure.}
 Intralayer exciton (coloured) can  efficiently emit light, while interlayer excitons (grey) show a much lower oscillator strength. However, driven by scattering with phonons (blue wave) their occupation can exceed the one of bright exciton by order of magnitude making them visible in low-temperature PL spectra. 
}
   \label{figure1}
\end{figure}

After optical excitation of an intralayer exciton in the TMD  layer, efficient  charge  transfer enables the hole to tunnel from the TMD to the organic layer, while still being attracted to the TMD electron (cf. Fig. \ref{figure1}). The spatially  separated  electron-hole  pair  is  an   interlayer exciton (also known as charge transfer or hybrid exciton). Due to their spatial separation,  ILXs in van der Waals heterostructures have a small oscillator strength and are thus in general darker than intralayer excitons (illustrated with the grey ILX in Fig. \ref{figure1}). While there is increasing number of studies on ILX signatures in PL spectra in TMD bilayers \cite{ovesen2019interlayer, merkl2019ultrafast}, little is known about the intra- and interlayer exciton landscape and their optical spectra characterizing TMD/molecule heterostructures. 
In this work, we undertake a join theory-experiment study to uncover the temperature dependent excitonic signatures of  tungsten disulfide (WS$_2$)/Tc heterostructure as an exemplary system for the broad class of TMD/molecule heterostructures.

To obtain microscopic insights, we apply a quantum-mechanical approach based on the density matrix formalism \cite{Kochbuch, Kira2006}. In particular, we solve the Wannier equation allowing us to microscopically resolve the entire intra- and interlayer exciton landscape in the investigated materials including binding energies and wave functions. Solving the generalized Elliott formula we also have microscopic access to their optical signatures \cite{brem2019phonon,selig2018dark}. Experimentally, we perform temperature-dependent photoluminescence (PL) measurements on WS$_2$/Tc heterostructures and demonstrate the appearance of ILXs at energies well below the optically excited intralayer exciton. In excellent agreement with theoretical predictions, we show how the ILX intensity significantly increases at lower temperatures. We find that these signatures are composed of  two close-lying ILXs originating from  two distinct electronic valleys in the TMD layer, reflecting the complex ILX landscape of TMDs.  Furthermore, we  demonstrate both in experiment and theory signatures of ILX phonon sidebands. The latter are identified as a low-energy shoulder of the ILX resonance.   Overall, our work provides new microscopic insights into the intriguing ILX landscape and the resulting optical response in technologically promising organic/inorganic heterostructures.\\

\begin{figure}[t!]
\includegraphics[width=\linewidth]{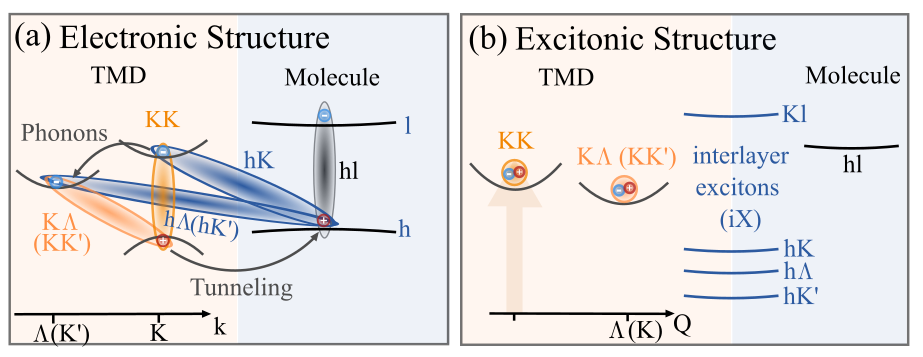} 
\caption{\textbf{Bandstructure of intra- and interlayer excitons.}
    (a) Electronic dispersion of a WS$_2$ monolayer around the high symmetry K and $\Lambda$ (K') points and the homo-lumo gap in tetracene (Tc) molecules. 
    (b) Excitonic dispersion  of the WS$_2$/Tc heterostructure including  intra- and interlayer states obtained from the solution of the Wannier equation. 
    }
   \label{figure2}
\end{figure}

{\noindent \textbf{Interlayer exciton landscape. }} 
We focus on heterostructures comprising of a  WS$_2$ monolayer and a thin crystalline film of Tc molecules. The latter attach to the surface of the TMD via van der Waals bonds and form a molecular lattice, as has been shown in previous studies \cite{park2018hybrid,amsterdam2020tailoring}. Tetracene molecules tend to form periodic lattices rather than distribute randomly on the surface \cite{rotter2016coupling,witte2004growth,el2010rubrene,kafer2007evolution,niederhausen2020tetracene}. Motivated by these findings, we describe the Tc layer as a quasi two-dimensional structure characterized by a homo-lumo gap. We take into account one state in the conduction and valence band denoted as l (lowest unoccupied molecular orbital) and h (highest occupied molecular orbital). This is justified since the bandwidth of the l and h bands in the molecular layer is very small compared to the TMD conduction and valence bands \cite{breuer2016structure,gavrila2004energy, fedorov2017first}. Hence, the number of observed ILXs will be primarily determined by the TMD band structure cf. Fig. \ref{figure2}. The molecular states, while slightly dispersive, have a quasi-flat dispersion, cf. Fig. \ref{figure2} owing to the relatively weak hopping between molecules \cite{fedorov2017first, cocchi2018polarized, doi2005ab}.

To describe the TMD/Tc heterostructure on a microscopic level, we use the density matrix formalism  \cite{Kochbuch, Kira2006, gunnar_prb,selig2018dark,brem2018exciton}.
For the TMD, we take into account the band structure around the high-symmetry points in the reciprocal space within an effective mass approximation \cite{andor}. 
In particular, we consider the valence band maxima at the K point and conduction band minima at the K,$\Lambda,$ and K' valley, which are crucial in determining the exciton landscape \cite{malic2018dark}. 
Depending on the location of the electron and hole, we can distinguish different intralayer KK, K$\Lambda$ and KK' excitons (yellow/orange ovals in Fig. \ref{figure2}) and molecular excitons hl (black oval in Fig. \ref{figure2}). Furthermore, there are spatially separated interlayer excitons hK, h$\Lambda$, hK', Kl  combining TMD and molecule states (blue ovals in Fig. \ref{figure2}). 
Here, the first/second letter describes the position of the hole/electron, e.g. the state hK corresponds to the hole located in the h state of the molecule, while the electron can be found in the K valley of the TMD layer, cf. Fig. \ref{figure2}(a). The quasi-flatness of the molecular dispersion means that ILXs are spread over a larger area of momentum-space, reflecting its real space localisation. Furthermore, since the Brillouin zones of WS$_2$ and Tc layers are considerably different, we assume that all ILXs are bright as no momentum transfer is required \cite{hagel2021}. 

As excitonic effects are crucial in TMDs \cite{Chernikov2014, gunnar_prb,arora2015excitonic,erminnpj}, we  project our equations into an excitonic basis with eigenenergies $\epsilon_\mu$ and wave functions $\varphi_q^\mu$. Note that the index $\mu$ can include both intralayer $\mu$ =(KK,K$\Lambda$,KK',hl) and interlayer states $\mu$ =(hK,h$\Lambda$,hK',Kl).  The excitonic eigenfunctions and eigenenergies are obtained by solving the Wannier equation \cite{Kochbuch, Kira2006,gunnar_prb}. 
 The Coulomb matrix element is calculated using an effective 2D Coulomb potential $V_{k}^\mu =  \frac{e_0^2}{k\epsilon_0\epsilon (k)} $ 
 with the dielectric function $\epsilon(k)$ \cite{ovesen2019interlayer}, depending on the screening of the two layers $\epsilon^{\text{TMD}}/ \epsilon^{\text{Tc}}$, the screening of the surrounding material $\epsilon^{\text{sur}}$, the thickness of  the layers $d^{\text{TMD}}/d^{\text{Tc}}$ and the interlayer distance $d^{\text{TMD-TC}}$. Table S1 in the supplementary summarizes the parameters used in our study. 
 The assumed thickness of the molecular film corresponds to a single layer of molecules in an upright configuration. Note that in experimental setups, the molecular layer of $\sim 20$ nm consists of more then one layer, however we assume that the main interaction and charge transfer takes place within the first layer. This is further justified since in ab-initio calculations, exciton wavefunctions in organic semiconductors exist primarily within single layer planes \cite{cocchi2018polarized}. 
Solving the Wannier equation, we obtain the binding energies of the entire exciton landscape including bright and dark intra- and interlayer excitons, as illustrated in Figure \ref{figure2}(b). While momentum-dark and bright intralayer excitons are easily defined, the precise nature of ILXs is harder to pin down. Changes in the growth conditions and substrate quality will change the structure of the molecular lattice and hence its dispersion. 

For WS$_2$/Tc heterostructures on a Si/SiO$_2$ substrate, we find an exciton binding energy of 196 meV for the intralayer exciton within the WS$_2$ layer. Furthermore, we find binding energies of 243 meV and 272 meV for the two bright hK and Kl interlayer exciton states, respectively. Although  counter-intuitive at first glance, the larger binding energy in the ILX state comes from the larger effective mass of the hole in the molecular layer. The  ILX has a larger binding energy of 328 (272) meV reflecting the larger effective mass at the $\Lambda$ (K') point. The binding energies are slightly lower then for intralayer excitons in the WS$_2$ monolayer due to the larger dielectric screening within a heterostructure. The energies are in a similar range as values reported for TMD/TMD heterostructures \cite{ovesen2019interlayer,brem2020hybridized, merkl2019ultrafast}.  While the large effective masses and smaller dielectric screening  in the molecular film lead to heavier excitons and hence enlarged binding energies, the relatively large thickness of the molecular layer compared to the TMD monolayer reduces the exciton binding energy.
 Note that the energies discussed so far are pure  binding energies $E^b_{\mu}$. For the absolute spectral position $E_{\mu}$ of excitons (shown in Fig. \ref{figure2}(b)), one needs  to take into account the band gap energy $E^g_{\mu}$ with $E_{\mu} = E^g_{\mu}-E^b_{\mu}$ . We shifted the whole exciton landscape to match the experimentally measured A exciton resonance at 2.0 eV. Overall, we find that the interlayer  hK' exciton is the energetically lowest state,  followed by $E_{\text{h}\Lambda}$ and $E_{\text{hK}}$, cf. Fig. \ref{figure2}(b) and Table S1.  The interlayer Kl exciton is located 1.1 eV above the bright KK exciton and thus has no influence on PL spectra and will be neglected in the following. Moreover, we do not consider the intralayer molecule state hl (located at 2.56 eV), since our focus lies on  optical excitation resonant to the energetically lower intralayer TMD exciton.\\

{\noindent \textbf{Interlayer excitons in photoluminescence.}} 
After determining the exciton landscape in WS$_2$/Tc heterostructures, we focus now on their optical response in PL spectra. The important (fixed) input parameters (effective masses, band gap energies)  are obtained from DFT calculations for the electronic bandstructure \cite{andor,breuer2016structure}, cf. Table S1 in the supplementary.
We exploit the equation of motion for the photon-assisted polarization to obtain the PL intensity $ I_{\text{PL}}$ for intra- and interlayer excitons resulting in the Elliott formula   \cite{brem2019phonon}  \begin{equation}
 \label{PL}
  I^{}_{\text{PL}}(\omega)\propto \sum_{\mu =\text{ KK,hK}} \frac{|M_{ \mu}|^2 \gamma_{\mu} N_{\mu}}{ (E_{\mu}-\omega)^2+(\gamma_{\mu}+\Gamma_{\mu })^2},   
 \end{equation}
where the position of the resonances is given by the energy $E_{\mu}$, the resonance width by the radiative ($\gamma_{\mu}$) and the non-radiative phonon-assisted  ($\Gamma_{\mu}$) dephasing, and finally the oscillator strength by the exciton-photon matrix element $M_{\mu} = \delta_{Q,0}\sum_q  \varphi_q^{\mu}  M_q^{\mu} \hat{U}_\mu$. The latter is determined by the exciton wave function $\varphi_q^{\mu}$, the  momentum-dependent optical matrix element \cite{gunnar_prb} $M_q^{\mu}$ and the transformation matrix $\hat{U}_{\mu}$ which includes  the tunneling. 
The Kronecker-delta ensures that only bright excitons within the light cone with $Q=0$ are optically active \cite{brem2019phonon}. 
The appearing exciton occupation $N_{\mu}$ is crucial in determining the spectral weight of different transitions in the PL spectrum. In this study, we assume thermalized Boltzmann distributions, since we discuss the stationary  PL only \cite{selig2018dark,brem2019phonon}. More details on the origin of Eq. \ref{PL} can be found in the supplementary.

\begin{figure}[t!]
\includegraphics[width=\linewidth]{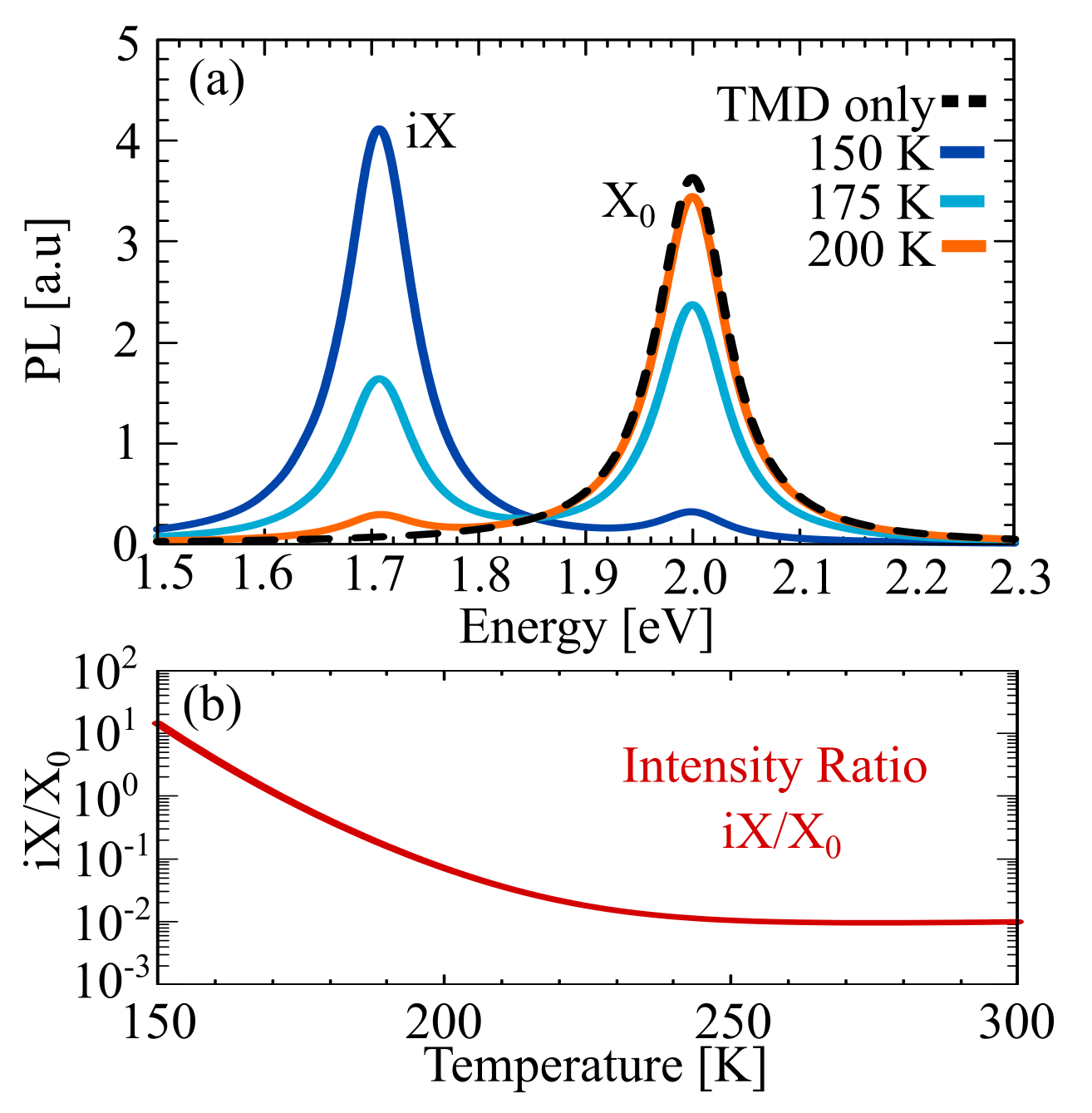} 
    \caption{\textbf{Optical signature of interlayer excitons.} 
         (a) Calculated PL spectrum of a WS$_2$/Tc heterostructure  on a Si-SiO$_2$ illustrating the drastically growing intensity of the hK' interlayer exciton resonance for reduced temperatures. For comparison, the calculated PL for a WS$_2$ monolayer at 200 K is shown with the dashed line exhibiting only the intralayer exciton resonance. (b)  The intensity ratio between the hK'  interlayer (iX) and KK intralayer excitons (X$_0$), plotted on a log scale. 
    }
   \label{figure3}
\end{figure}

Evaluating Eq. (\ref{PL}) numerically, we investigate  optical signatures of intra- and interlayer excitons in the WS$_2$/Tc heterostructure. The PL spectrum at large temperature ($>$200 K) exhibits a pronounced resonance at 2 eV stemming from the bright intralayer KK exciton, cf. Fig. \ref{figure3}(a). As the temperature decreases we observe the emergence of a second peak around 1.71 eV that corresponds to the expected position of the ILX. Further lowering the temperature increases the ratio between the inter- and intralayer exciton peaks (cf. Fig \ref{figure3} (b)) until at low temperatures ($<$150 K) the intralayer contributions becomes negligible and the PL spectrum is entirely dominated by ILXs. The lower the temperature, the higher the occupation of the energetically lower-lying ILX states which leads to a more intense ILX resonance in the PL spectra.  Therefore, the temperature acts as an externally accessible knob allowing the relative visibility of ILXs to be tuned.  Figure \ref{figure3} (b) shows the temperature-dependent intensity ratio of inter- and intralayer excitons on a logarithmic scale, which resembles the Boltzmann distribution.  The ILX in TMD/Tc heterostructures is visible at higher temperatures than in TMD bilayers. The main reason is that the  interlayer hK' exciton is about 20 meV lower in energy than the corresponding ILX states in TMD bilayers \cite{ovesen2019interlayer,brem2020hybridized,rivera2015observation,hong2014ultrafast,kunstmann2018momentum}. This leads to a higher occupation $N_\mu$ in equilibrium and thus to the stronger ILX contribution, cf. Eq. (\ref{PL}).
 Note that we have neglected temperature-induced spectral shifts, linewidth changes and possible modifications in the molecular layer, since they are not expected to have an impact on the  qualitative temperature trend of PL intensity ratios.\\


{\noindent \textbf{Signatures of ILX phonon-sidebands.}} So far, we have focused on direct ILX emission, however  we know from TMD monolayers that  intralayer excitons show pronounced signatures via phonon-assisted indirect emission resulting in phonon sidebands \cite{brem2019phonon,feldtmann2009phonon,chernikov2012phonon}.  In TMD homo- and heterobilayers, indirect, phonon-assisted signatures of  dark interlayer excitons have been observed \cite{kunstmann2018momentum, brem2020hybridized, hagel2021}. 
To account for these indirect emission channels also for ILX states, we exploit a generalized Elliott formula \cite{brem2019phonon} 
\begin{equation}
    \label{PLdark}
I_{\text{PL}}^{} (\omega)\propto \sum_{\substack{
\nu \mu  \\
 Q,\alpha\pm}} 
 \frac{\Theta^{\nu} (\omega)
   |D^{\nu \mu }_{Q\alpha}|^2 \Gamma_{\mu}  N_{\mu} \eta_\alpha^{\pm} }{(E_{\mu}\pm\Omega^\alpha_Q -\omega)^2 +(\Gamma_{\mu})^2}
\end{equation}
with $\Theta^{\nu  }(\omega)=\frac{|M^{\nu}|^2 }{(E_{\nu }-\omega)^2+(\gamma^{\nu}+\Gamma^{\nu})^2}$. We take into account all ILX states $\nu  = (\text{KK, K}\Lambda, \text{KK}^\prime, \text{hK, h}\Lambda,\text{hK'})$.   
The position of resonances is  determined by the energy of the corresponding exciton $E_{\mu}$ plus/minus the energy of the absorbed/emitted phonon $\pm\Omega^{\alpha}_{Q}$ in the mode $\alpha$. We account for longitudinal and transversal optical and acoustic phonon modes.   The temperature dependent phonon occupation $n^{\text{ph}}_{\alpha}$ enters in  $\eta_\alpha^{\pm} = \left(
\frac{1}{2} \mp \frac{1}{2} + n^{\text{ph}}_{\alpha}
\right)$  corresponding  to  a Bose  distribution (bath approximation \cite{axt1996influence}).
In contrast to direct emission, phonon-sidebands cannot decay radiatively \cite{selig2016excitonic}, hence their spectral width is only determined by non-radiative dephasing  $\Gamma_{\mu}$ \cite{brem2019phonon,selig2016excitonic,selig2018dark}.
The oscillator strength of phonon-assisted indirect emission scales with the exciton-phonon scattering element  $|D^{\nu\mu}_{Q \alpha}|^2 = \sum_q \varphi_q^\mu g^{\nu  \mu}_Q \varphi_{q+\alpha Q}^\nu $, where $g^{\nu  \mu}_Q$ is the electron-phonon coupling element  \cite{selig2018dark,brem2019phonon}.%

\begin{figure}[t!]
\includegraphics[width=\linewidth]{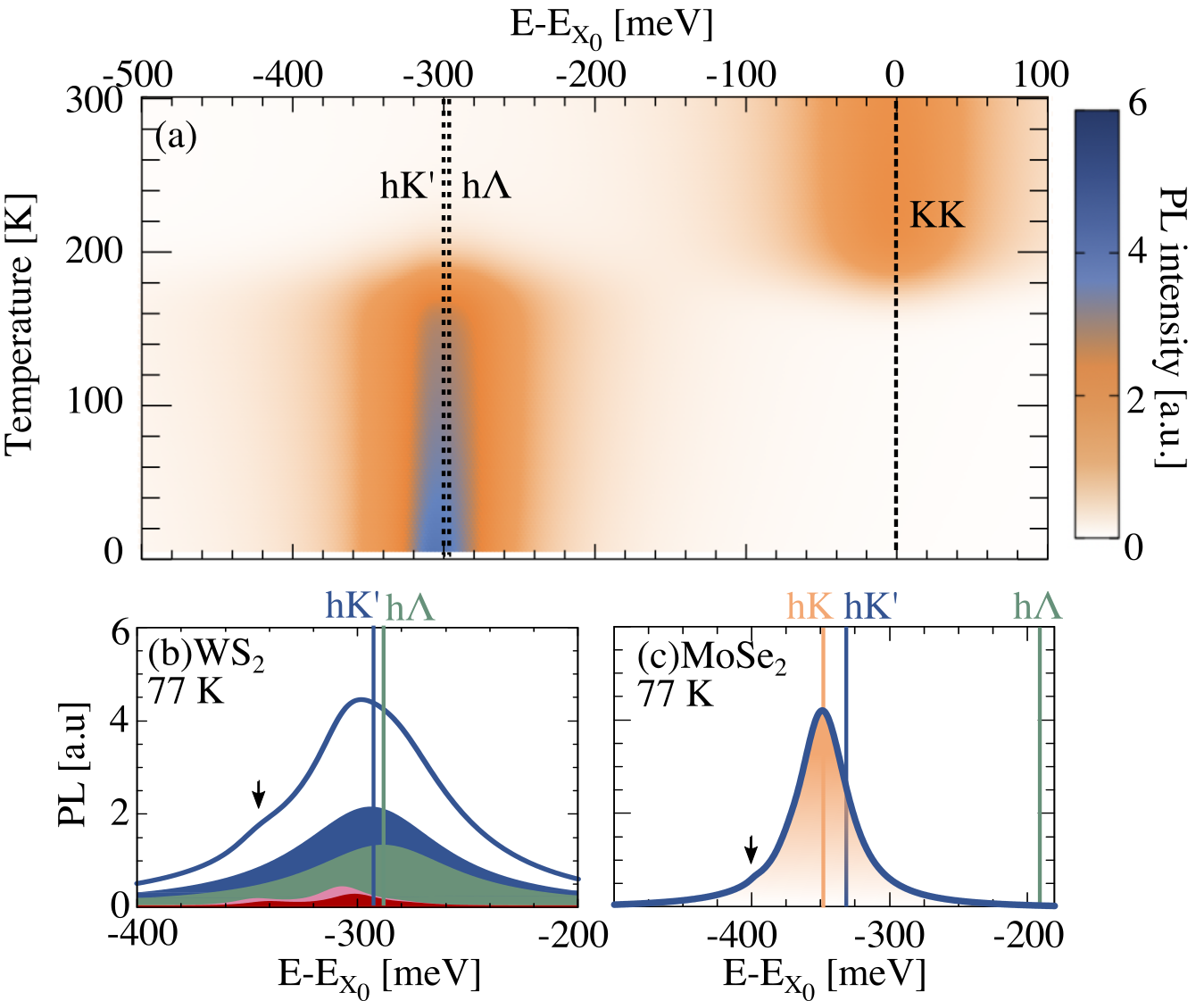} 
    \caption{\textbf{Fine structure of interlayer excitons.} (a) Calculated temperature-dependent PL of WS$_2$/Tc- heterostructure exhibiting signatures from intra- and interlayer excitons. Note that the origin of the energy axis has been shifted to the position of the bright KK exciton ($E_{X_0}$). (b)Constant temperature cut outlining the contributions of the phonon-assisted indirect PL from  interlayer exciton hK'  and h$\Lambda$. The zero phonon peak contribution is shown in shaded-blue (green), while the phonon-sideband contributions are pink (red) for the hK' (h$\Lambda$) excitons.  (c) For comparison, we also show the low-temperature PL for MoSe$_2$ exhibiting less significant phonon sidebands.}
   \label{figure4}
\end{figure}

Evaluating Eqs. (\ref{PL}) and (\ref{PLdark}), we calculate
the temperature-dependent PL including direct  and indirect phonon-assisted emission from excitons in the investigated WS$_2$/Tc heterostructure, cf. Fig. \ref{figure4}(a).  We find that at room temperature the PL is  dominated by the direct KK exciton $X_0$, while the  interlayer excitons (hK'/h$\Lambda$) located approximately 300 meV below $X_0$ becomes dominant in the temperature range below 200 K. At very low temperatures new resonances at lower energies appear and can be assigned to phonon-assisted emission from  interlayer h$\Lambda$ and hK' excitons, respectively.
To better understand these phonon sidebands, we zoom into  the low-energy ILX contribution at 77 K, cf. Fig. \ref{figure4}(b). We find that the energetically lowest  hK' interlayer exciton (blue-shaded) has the largest PL contribution. It consists of a direct zero-phonon peak at -295 and two smaller peaks (pink-shaded) at  -310  and -345 meV stemming from indirect light emission driven by acoustic and optical  phonons, respectively. Note that these peaks appear at the location $E_{\text{hK'}}\pm \Omega^{\text{A/O}}$, i.e. they are shifted by the phonon energy. We  find that the h$\Lambda$ interlayer exciton  has a smaller direct contribution (green-shaded) reflecting its lower occupation compared to hK' states as well as a smaller phonon-sideband contribution (red-shaded). By increasing the temperature, the ILX occupation  decreases, reducing their weight in PL spectra.   The total PL spectra (blue line) is therefore asymmetric with a clear optical sideband indicated by an arrow. Furthermore, the sidebands from the acoustic spectrum leads to a slight shift in the maximum PL intensity, about 10 meV below the PL resonance. Finally,  there is also a slight asymmetry at higher energies due to smaller h$\Lambda$ interlayer exciton contribution.

 For comparison,  in MoSe$_2$  the hK exciton is the energetically lowest state \cite{selig2018dark,malic2018dark,deilmann2019finite,madeo2020directly}. As a result, the low-temperature PL spectrum is qualitatively different exhibiting only one peak at -346 meV stemming from the  hK interlayer exciton, cf. Fig. \ref{figure4}(c). There are no signatures from the hK' and h$\Lambda$ interlayer exciton due to their lower occupation, however a small phonon-sideband originating from the hK exciton is visible in the spectra (marked by an arrow). \\



\begin{figure}[t!]
\includegraphics[width=0.99\linewidth]{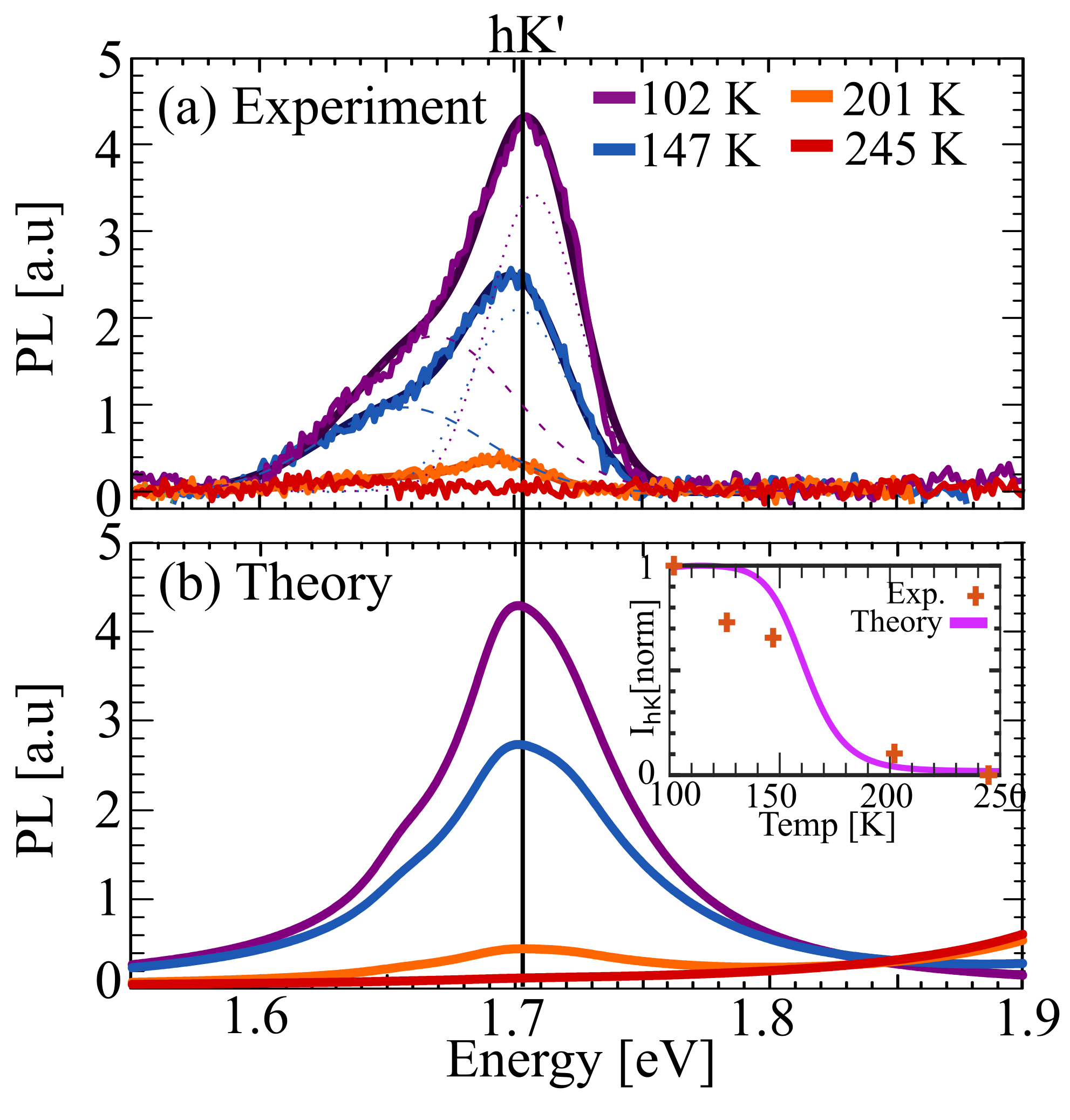} 
    \caption{\textbf{Experiment-theory comparison.} (a) Temperature-dependent PL spectra (a) measured in experiment and (b) predicted from microscopic theory  for WS$_2$/Tc on an SiO$_2$/Si substrate. Note that the theoretical spectra are shifted to align with experimental hK' exciton peak. Raw data can be fitted to using two Gaussian peaks (dotted and dashed line), which we attribute to the direct PL and phonon-sidebands, respectively. Inset shows a quantitative comparison of the evolution of the intensity of the hK' resonance as a function of temperature, (normalised to the peak maximum). In (a) and in the inset, we removed contributions from the background PL in experiment to allow for a more accurate comparison.}
   \label{figure5}
\end{figure}

{\noindent \textbf{Theory-experiment comparison.}}
To experimentally verify the theoretically predicted low temperature PL interlayer exciton signatures, we fabricated a high quality WS$_2$/Tc heterostructure and collected its PL spectra over a broad temperature range (100-245K). To prepare the heterostructure, a monolayer of WS$_2$ was exfoliated onto a silicon dioxide wafer (oxide thickness of 90nm) from bulk WS$_2$. A ~20nm layer of tetracene was then thermally evaporated onto the monolayer. Further information on sample preparation is included in the SI. We selectively excited WS$_2$ with 2.14 eV light to prevent the signal from interlayer excitons to become masked by Tc emission.
 Fig. \ref{figure5}(a) shows the measured temperature dependent PL of the heterostructure.  At 245K, the  interlayer hK' exciton emits at 1.71eV and its intensity significantly grows with decreasing temperature. This reflects the increasing occupation of the hK' excitonic state, which is in excellent agreement with our calculations, cf.  Fig. \ref{figure5}(b). 
  In the inset, we compare the experimental (red crosses) and theoretical (pink curve) evolution of the ILX intensity as a function of  temperature.  We normalize the ILX intensity to the value at 100 K. We find a good  agreement between our theoretical model and experimental data, with a significant PL from the hK' exciton in the temperature range of  100-200 K.    By performing a double-Gaussian fit, and taking the area under the curves,  we are able to attribute the temperature evolution of the ILX peak to a change in exction occupation $N_\mu$, rather than a broadening effect (see SI).
 Note that while we have the means to microscopically estimate the radiative and phonon-induced broadening of exciton peaks, broadening mechanisms in TMD-molecule heterostructures are currently not well understood, and our model underestimates the observed experimental broadening. Therefore, to better compare to experiment we extract values for the broadening from the experimental data and use a phenomenological model \cite{brem2020hybridized} to describe the temperature dependence in Fig. \ref{figure5}(b).

At around 100 K, we observe the increase in the intensity of a shoulder peak around 1.67 eV as indicated by the Gaussian fit (Fig. 5a).
This signature also appears in the theoretically calculated PL spectrum, cf. Fig \ref{figure5}(b) at 1.66 eV, and can be ascribed to a phonon sideband stemming from the indirect, phonon-assisted recombination of the hK' interlayer exciton states. In the theory, the sideband is red-shifted with respect to the position of the hK' resonance at 
1.70 eV, reflecting the emission of an optical phonon  \cite{selig2018dark}. At low temperatures, the signature of the phonon sideband increases relative to the zero-phonon peak due to the lower phonon-induced broadening.  At temperatures above 200 K, this phonon sideband vanishes and is hardly visible in both the theory and experiment due to the lower occupation of the hK' exciton and increased broadening. This experimental feature is consistent for different trials (see SI).
 The spectral profiles including peak positions and temperature dependence are all typical hallmarks of interlayer phonon-sidebands \cite{doi:10.1021/acsphotonics.0c00866}. Furthermore, we have performed lifetime measurements at 102 K and 245 K, cf. the SI. Upon cooling to 102 K, the heterostructure’s lifetime is considerably enhanced. The large increase in lifetime provides further evidence that at low temperature we indeed access ILX states.

While we obtain an excellent qualitative agreement between theory and experiment, some quantitative discrepancies do exist.  In particular, there is  an increasing PL signal at higher energies above the hK' resonance in the experiment, while in theory the PL is strongly suppressed in this spectral region. This could be traced back to a particularly broad intralayer exciton resonance in the experiment that is expected to occur at about 2 eV (see supplementary).

Interestingly, the peak position of the hK' interlayer exciton does not spectrally shift in the experiment. This suggests that the temperature-dependent Varshni energy shifts, arising from the thermal expansion of the crystal lattice \cite{varshni1967temperature}, are negligible for interlayer excitons, justifying the omission of such effects in our theory. The suppressed Varshni shift could be explained by the involvement of the molecular lattice, which is expected to be less susceptible to thermal expansion when placed on a substrate such as WS$_2$. Finally, the position of the observed phonon-sideband is slightly lower in energy than theoretically predicted (cf. arrows in Fig. \ref{figure5}). While we have fairly accurate estimates of the phonon dispersion for a pure TMD, phonon energies in a WS$_2$/Tc heterostructure might differ. A first principle calculation of the phonon dispersion for the TMD/Tc heterostructure is beyond the scope of this work.

In conclusion, we have presented a joint theory-experiment study on the optical response of interlayer excitons in a TMD/tetracene heterostructure. We microscopically reveal the entire landscape of interlayer excitons living in both molecule and TMD layer and show their signatures in temperature-dependent PL spectra. In particular, we find both in experiment and theory a clear signature of  interlayer excitons in PL spectra and show how their intensity  can be boosted by variation in  temperature. Furthermore, we show the appearance of an additional low-energy PL signature at temperatures around 100 K  stemming from indirect phonon-assisted emission from interlayer exciton states. 
Our work provides new insights into the microscopic nature of interlayer excitons in TMD/molecule heterostructures  and could guide future studies in the growing research field of technologically promising organic/inorganic heterostructures.\\

{\noindent \textbf{Acknowledgments:}} 
This project has received funding from Deutsche Forschungsgemeinschaft (DFG, German Research Foundation)  through SFB 1083 and the European Union's Horizon
2020 research and innovation program under grant agreement No 881603 (Graphene Flagship). The experimental work at Purdue was supported by the US Department of Energy, Office of Basic Energy Science, through award DE-SC0016356 (optical spectroscopy) and DE-SC0022082 (sample fabrication).

\bibliographystyle{achemso}

\begin{mcitethebibliography}{50}
\providecommand*\natexlab[1]{#1}
\providecommand*\mciteSetBstSublistMode[1]{}
\providecommand*\mciteSetBstMaxWidthForm[2]{}
\providecommand*\mciteBstWouldAddEndPuncttrue
  {\def\EndOfBibitem{\unskip.}}
\providecommand*\mciteBstWouldAddEndPunctfalse
  {\let\EndOfBibitem\relax}
\providecommand*\mciteSetBstMidEndSepPunct[3]{}
\providecommand*\mciteSetBstSublistLabelBeginEnd[3]{}
\providecommand*\EndOfBibitem{}
\mciteSetBstSublistMode{f}
\mciteSetBstMaxWidthForm{subitem}{(\alph{mcitesubitemcount})}
\mciteSetBstSublistLabelBeginEnd
  {\mcitemaxwidthsubitemform\space}
  {\relax}
  {\relax}

\bibitem[Geim and Grigorieva(2013)Geim, and Grigorieva]{geim2013van}
Geim,~A.~K.; Grigorieva,~I.~V. Van der Waals heterostructures. \emph{Nature}
  \textbf{2013}, \emph{499}, 419--425\relax
\mciteBstWouldAddEndPuncttrue
\mciteSetBstMidEndSepPunct{\mcitedefaultmidpunct}
{\mcitedefaultendpunct}{\mcitedefaultseppunct}\relax
\EndOfBibitem
\bibitem[Liu \latin{et~al.}(2019)Liu, Zhang, He, Wang, and Liu]{liu2019recent}
Liu,~Y.; Zhang,~S.; He,~J.; Wang,~Z.~M.; Liu,~Z. Recent progress in the
  fabrication, properties, and devices of heterostructures based on 2D
  materials. \emph{Nano-Micro Letters} \textbf{2019}, \emph{11}, 1--24\relax
\mciteBstWouldAddEndPuncttrue
\mciteSetBstMidEndSepPunct{\mcitedefaultmidpunct}
{\mcitedefaultendpunct}{\mcitedefaultseppunct}\relax
\EndOfBibitem
\bibitem[Alexeev \latin{et~al.}(2019)Alexeev, Ruiz-Tijerina, Danovich, Hamer,
  Terry, Nayak, Ahn, Pak, Lee, Sohn, Molas, Koperski, Watanabe, Taniguchi,
  Novoselov, Gorbachev, Shin, Fal’ko, and
  Tartakovskii]{alexeev2019resonantly}
Alexeev,~E.~M. \latin{et~al.}  Resonantly hybridized excitons in moir{\'e}
  superlattices in van der Waals heterostructures. \emph{Nature} \textbf{2019},
  \emph{567}, 81--86\relax
\mciteBstWouldAddEndPuncttrue
\mciteSetBstMidEndSepPunct{\mcitedefaultmidpunct}
{\mcitedefaultendpunct}{\mcitedefaultseppunct}\relax
\EndOfBibitem
\bibitem[Jin \latin{et~al.}(2019)Jin, Regan, Yan, Utama, Wang, Zhao, Qin, Yang,
  Zheng, Shi, Watanabe, Taniguchi, Tongay, Zettl, and Wang]{jin2019observation}
Jin,~C.; Regan,~E.~C.; Yan,~A.; Utama,~M. I.~B.; Wang,~D.; Zhao,~S.; Qin,~Y.;
  Yang,~S.; Zheng,~Z.; Shi,~S.; Watanabe,~K.; Taniguchi,~T.; Tongay,~S.;
  Zettl,~A.; Wang,~F. Observation of moir{\'e} excitons in WSe 2/WS 2
  heterostructure superlattices. \emph{Nature} \textbf{2019}, \emph{567},
  76--80\relax
\mciteBstWouldAddEndPuncttrue
\mciteSetBstMidEndSepPunct{\mcitedefaultmidpunct}
{\mcitedefaultendpunct}{\mcitedefaultseppunct}\relax
\EndOfBibitem
\bibitem[Tran \latin{et~al.}(2019)Tran, Moody, Wu, Lu, Choi, Kim, Rai, Sanchez,
  Quan, Singh, Embley, Zepeda, Campbell, Autry, Taniguchi, Watanabe, Lu,
  Banerjee, Silverman, Kim, Tutuc, Yang, MacDonald, and Li]{Tran2019EvidenceFM}
Tran,~K.~X. \latin{et~al.}  Evidence for moir{\'e} excitons in van der Waals
  heterostructures. \emph{Nature} \textbf{2019}, \emph{567}, 71--75\relax
\mciteBstWouldAddEndPuncttrue
\mciteSetBstMidEndSepPunct{\mcitedefaultmidpunct}
{\mcitedefaultendpunct}{\mcitedefaultseppunct}\relax
\EndOfBibitem
\bibitem[Merkl \latin{et~al.}(2019)Merkl, Mooshammer, Steinleitner, Girnghuber,
  Lin, Nagler, Holler, Sch{\"u}ller, Lupton, Korn, Ovesen, Brem, Ermin, and
  Huber]{merkl2019ultrafast}
Merkl,~P.; Mooshammer,~F.; Steinleitner,~P.; Girnghuber,~A.; Lin,~K.-Q.;
  Nagler,~P.; Holler,~J.; Sch{\"u}ller,~C.; Lupton,~J.~M.; Korn,~T.;
  Ovesen,~S.; Brem,~S.; Ermin,~M.; Huber,~R. Ultrafast transition between
  exciton phases in van der Waals heterostructures. \emph{Nature materials}
  \textbf{2019}, \emph{18}, 691--696\relax
\mciteBstWouldAddEndPuncttrue
\mciteSetBstMidEndSepPunct{\mcitedefaultmidpunct}
{\mcitedefaultendpunct}{\mcitedefaultseppunct}\relax
\EndOfBibitem
\bibitem[Brem \latin{et~al.}(2020)Brem, Lin, Gillen, Bauer, Maultzsch, Lupton,
  and Malic]{brem2020hybridized}
Brem,~S.; Lin,~K.-Q.; Gillen,~R.; Bauer,~J.~M.; Maultzsch,~J.; Lupton,~J.~M.;
  Malic,~E. Hybridized intervalley moir{\'e} excitons and flat bands in twisted
  WSe 2 bilayers. \emph{Nanoscale} \textbf{2020}, \emph{12}, 11088--11094\relax
\mciteBstWouldAddEndPuncttrue
\mciteSetBstMidEndSepPunct{\mcitedefaultmidpunct}
{\mcitedefaultendpunct}{\mcitedefaultseppunct}\relax
\EndOfBibitem
\bibitem[Brem \latin{et~al.}(2020)Brem, Linder{\"a}lv, Erhart, and
  Malic]{brem2020tunable}
Brem,~S.; Linder{\"a}lv,~C.; Erhart,~P.; Malic,~E. Tunable Phases of Moir{\'e}
  Excitons in van der Waals Heterostructures. \emph{Nano letters}
  \textbf{2020}, \emph{20}, 8534--8540\relax
\mciteBstWouldAddEndPuncttrue
\mciteSetBstMidEndSepPunct{\mcitedefaultmidpunct}
{\mcitedefaultendpunct}{\mcitedefaultseppunct}\relax
\EndOfBibitem
\bibitem[Deilmann and Thygesen(2019)Deilmann, and Thygesen]{deilmann2019finite}
Deilmann,~T.; Thygesen,~K.~S. Finite-momentum exciton landscape in mono-and
  bilayer transition metal dichalcogenides. \emph{2D Materials} \textbf{2019},
  \emph{6}, 035003\relax
\mciteBstWouldAddEndPuncttrue
\mciteSetBstMidEndSepPunct{\mcitedefaultmidpunct}
{\mcitedefaultendpunct}{\mcitedefaultseppunct}\relax
\EndOfBibitem
\bibitem[Malic \latin{et~al.}(2018)Malic, Selig, Feierabend, Brem,
  Christiansen, Wendler, Knorr, and Bergh{\"a}user]{malic2018dark}
Malic,~E.; Selig,~M.; Feierabend,~M.; Brem,~S.; Christiansen,~D.; Wendler,~F.;
  Knorr,~A.; Bergh{\"a}user,~G. Dark excitons in transition metal
  dichalcogenides. \emph{Physical Review Materials} \textbf{2018}, \emph{2},
  014002\relax
\mciteBstWouldAddEndPuncttrue
\mciteSetBstMidEndSepPunct{\mcitedefaultmidpunct}
{\mcitedefaultendpunct}{\mcitedefaultseppunct}\relax
\EndOfBibitem
\bibitem[Mad{\'e}o \latin{et~al.}(2020)Mad{\'e}o, Man, Sahoo, Campbell, Pareek,
  Wong, Al-Mahboob, Chan, Karmakar, Mariserla, \latin{et~al.}
  others]{madeo2020directly}
Mad{\'e}o,~J.; Man,~M.~K.; Sahoo,~C.; Campbell,~M.; Pareek,~V.; Wong,~E.~L.;
  Al-Mahboob,~A.; Chan,~N.~S.; Karmakar,~A.; Mariserla,~B. M.~K.,
  \latin{et~al.}  Directly visualizing the momentum-forbidden dark excitons and
  their dynamics in atomically thin semiconductors. \emph{Science}
  \textbf{2020}, \emph{370}, 1199--1204\relax
\mciteBstWouldAddEndPuncttrue
\mciteSetBstMidEndSepPunct{\mcitedefaultmidpunct}
{\mcitedefaultendpunct}{\mcitedefaultseppunct}\relax
\EndOfBibitem
\bibitem[Wallauer \latin{et~al.}(2021)Wallauer, Perea-Causin, M{\"u}nster,
  Zajusch, Brem, G{\"u}dde, Tanimura, Lin, Huber, Malic, \latin{et~al.}
  others]{wallauer2021momentum}
Wallauer,~R.; Perea-Causin,~R.; M{\"u}nster,~L.; Zajusch,~S.; Brem,~S.;
  G{\"u}dde,~J.; Tanimura,~K.; Lin,~K.-Q.; Huber,~R.; Malic,~E., \latin{et~al.}
  Momentum-Resolved Observation of Exciton Formation Dynamics in Monolayer WS2.
  \emph{Nano Letters} \textbf{2021}, \emph{21}, 5867--5873\relax
\mciteBstWouldAddEndPuncttrue
\mciteSetBstMidEndSepPunct{\mcitedefaultmidpunct}
{\mcitedefaultendpunct}{\mcitedefaultseppunct}\relax
\EndOfBibitem
\bibitem[Rivera \latin{et~al.}(2015)Rivera, Schaibley, Jones, Ross, Wu,
  Aivazian, Klement, Seyler, Clark, Ghimire, \latin{et~al.}
  others]{rivera2015observation}
Rivera,~P.; Schaibley,~J.~R.; Jones,~A.~M.; Ross,~J.~S.; Wu,~S.; Aivazian,~G.;
  Klement,~P.; Seyler,~K.; Clark,~G.; Ghimire,~N.~J., \latin{et~al.}
  Observation of long-lived interlayer excitons in monolayer MoSe 2--WSe 2
  heterostructures. \emph{Nature communications} \textbf{2015}, \emph{6},
  1--6\relax
\mciteBstWouldAddEndPuncttrue
\mciteSetBstMidEndSepPunct{\mcitedefaultmidpunct}
{\mcitedefaultendpunct}{\mcitedefaultseppunct}\relax
\EndOfBibitem
\bibitem[Hong \latin{et~al.}(2014)Hong, Kim, Shi, Zhang, Jin, Sun, Tongay, Wu,
  Zhang, and Wang]{hong2014ultrafast}
Hong,~X.; Kim,~J.; Shi,~S.-F.; Zhang,~Y.; Jin,~C.; Sun,~Y.; Tongay,~S.; Wu,~J.;
  Zhang,~Y.; Wang,~F. Ultrafast charge transfer in atomically thin MoS 2/WS 2
  heterostructures. \emph{Nature nanotechnology} \textbf{2014}, \emph{9},
  682\relax
\mciteBstWouldAddEndPuncttrue
\mciteSetBstMidEndSepPunct{\mcitedefaultmidpunct}
{\mcitedefaultendpunct}{\mcitedefaultseppunct}\relax
\EndOfBibitem
\bibitem[Kunstmann \latin{et~al.}(2018)Kunstmann, Mooshammer, Nagler, Chaves,
  Stein, Paradiso, Plechinger, Strunk, Sch{\"u}ller, Seifert, \latin{et~al.}
  others]{kunstmann2018momentum}
Kunstmann,~J.; Mooshammer,~F.; Nagler,~P.; Chaves,~A.; Stein,~F.; Paradiso,~N.;
  Plechinger,~G.; Strunk,~C.; Sch{\"u}ller,~C.; Seifert,~G., \latin{et~al.}
  Momentum-space indirect interlayer excitons in transition-metal
  dichalcogenide van der Waals heterostructures. \emph{Nature Physics}
  \textbf{2018}, \emph{14}, 801--805\relax
\mciteBstWouldAddEndPuncttrue
\mciteSetBstMidEndSepPunct{\mcitedefaultmidpunct}
{\mcitedefaultendpunct}{\mcitedefaultseppunct}\relax
\EndOfBibitem
\bibitem[Zhu \latin{et~al.}(2018)Zhu, Yuan, Zhao, Zhou, Wan, Mei, and
  Huang]{zhu2018highly}
Zhu,~T.; Yuan,~L.; Zhao,~Y.; Zhou,~M.; Wan,~Y.; Mei,~J.; Huang,~L. Highly
  mobile charge-transfer excitons in two-dimensional WS2/tetracene
  heterostructures. \emph{Science Advances} \textbf{2018}, \emph{4},
  eaao3104\relax
\mciteBstWouldAddEndPuncttrue
\mciteSetBstMidEndSepPunct{\mcitedefaultmidpunct}
{\mcitedefaultendpunct}{\mcitedefaultseppunct}\relax
\EndOfBibitem
\bibitem[Li \latin{et~al.}(2020)Li, Dong, Zhang, Li, Wang, Wang, Zhang, and
  Zhang]{li2020recent}
Li,~H.; Dong,~Z.; Zhang,~Y.; Li,~L.; Wang,~Z.; Wang,~C.; Zhang,~K.; Zhang,~H.
  Recent progress and strategies in photodetectors based on 2D
  inorganic/organic heterostructures. \emph{2D Materials} \textbf{2020},
  \emph{8}, 012001\relax
\mciteBstWouldAddEndPuncttrue
\mciteSetBstMidEndSepPunct{\mcitedefaultmidpunct}
{\mcitedefaultendpunct}{\mcitedefaultseppunct}\relax
\EndOfBibitem
\bibitem[Gobbi \latin{et~al.}(2018)Gobbi, Orgiu, and Samor{\`\i}]{gobbi20182d}
Gobbi,~M.; Orgiu,~E.; Samor{\`\i},~P. When 2D materials meet molecules:
  opportunities and challenges of hybrid organic/inorganic van der Waals
  heterostructures. \emph{Advanced Materials} \textbf{2018}, \emph{30},
  1706103\relax
\mciteBstWouldAddEndPuncttrue
\mciteSetBstMidEndSepPunct{\mcitedefaultmidpunct}
{\mcitedefaultendpunct}{\mcitedefaultseppunct}\relax
\EndOfBibitem
\bibitem[Lee \latin{et~al.}(2014)Lee, Lee, Van Der~Zande, Han, Cui, Arefe,
  Nuckolls, Heinz, Hone, and Kim]{lee2014heterostructures}
Lee,~G.-H.; Lee,~C.-H.; Van Der~Zande,~A.~M.; Han,~M.; Cui,~X.; Arefe,~G.;
  Nuckolls,~C.; Heinz,~T.~F.; Hone,~J.; Kim,~P. Heterostructures based on
  inorganic and organic van der Waals systems. \emph{Apl Materials}
  \textbf{2014}, \emph{2}, 092511\relax
\mciteBstWouldAddEndPuncttrue
\mciteSetBstMidEndSepPunct{\mcitedefaultmidpunct}
{\mcitedefaultendpunct}{\mcitedefaultseppunct}\relax
\EndOfBibitem
\bibitem[Costa \latin{et~al.}(2016)Costa, Taveira, Lima, Mendes, and
  Santos]{costa2016optical}
Costa,~J.~C.; Taveira,~R.~J.; Lima,~C.~F.; Mendes,~A.; Santos,~L.~M. Optical
  band gaps of organic semiconductor materials. \emph{Optical Materials}
  \textbf{2016}, \emph{58}, 51--60\relax
\mciteBstWouldAddEndPuncttrue
\mciteSetBstMidEndSepPunct{\mcitedefaultmidpunct}
{\mcitedefaultendpunct}{\mcitedefaultseppunct}\relax
\EndOfBibitem
\bibitem[Ovesen \latin{et~al.}(2019)Ovesen, Brem, Linder{\"a}lv, Kuisma, Korn,
  Erhart, Selig, and Malic]{ovesen2019interlayer}
Ovesen,~S.; Brem,~S.; Linder{\"a}lv,~C.; Kuisma,~M.; Korn,~T.; Erhart,~P.;
  Selig,~M.; Malic,~E. Interlayer exciton dynamics in van der Waals
  heterostructures. \emph{Communications Physics} \textbf{2019}, \emph{2},
  1--8\relax
\mciteBstWouldAddEndPuncttrue
\mciteSetBstMidEndSepPunct{\mcitedefaultmidpunct}
{\mcitedefaultendpunct}{\mcitedefaultseppunct}\relax
\EndOfBibitem
\bibitem[Haug and Koch(2009)Haug, and Koch]{Kochbuch}
Haug,~H.; Koch,~S.~W. \emph{Quantum theory of the optical and electronic
  properties of semiconductors}; World Scientific Publishing Company,
  2009\relax
\mciteBstWouldAddEndPuncttrue
\mciteSetBstMidEndSepPunct{\mcitedefaultmidpunct}
{\mcitedefaultendpunct}{\mcitedefaultseppunct}\relax
\EndOfBibitem
\bibitem[Kira and Koch(2006)Kira, and Koch]{Kira2006}
Kira,~M.; Koch,~S. Many-body correlations and excitonic effects in
  semiconductor spectroscopy. \emph{Progress in Quantum Electronics}
  \textbf{2006}, \emph{30}, 155 -- 296\relax
\mciteBstWouldAddEndPuncttrue
\mciteSetBstMidEndSepPunct{\mcitedefaultmidpunct}
{\mcitedefaultendpunct}{\mcitedefaultseppunct}\relax
\EndOfBibitem
\bibitem[Brem \latin{et~al.}(2020)Brem, Ekman, Christiansen, Katsch, Selig,
  Robert, Marie, Urbaszek, Knorr, and Malic]{brem2019phonon}
Brem,~S.; Ekman,~A.; Christiansen,~D.; Katsch,~F.; Selig,~M.; Robert,~C.;
  Marie,~X.; Urbaszek,~B.; Knorr,~A.; Malic,~E. Phonon-assisted
  photoluminescence from indirect excitons in monolayers of transition-metal
  dichalcogenides. \emph{Nano Letters} \textbf{2020}, \emph{20},
  2849--2856\relax
\mciteBstWouldAddEndPuncttrue
\mciteSetBstMidEndSepPunct{\mcitedefaultmidpunct}
{\mcitedefaultendpunct}{\mcitedefaultseppunct}\relax
\EndOfBibitem
\bibitem[Selig \latin{et~al.}(2018)Selig, Bergh{\"a}user, Richter,
  Bratschitsch, Knorr, and Malic]{selig2018dark}
Selig,~M.; Bergh{\"a}user,~G.; Richter,~M.; Bratschitsch,~R.; Knorr,~A.;
  Malic,~E. Dark and bright exciton formation, thermalization, and
  photoluminescence in monolayer transition metal dichalcogenides. \emph{2D
  Materials} \textbf{2018}, \emph{5}, 035017\relax
\mciteBstWouldAddEndPuncttrue
\mciteSetBstMidEndSepPunct{\mcitedefaultmidpunct}
{\mcitedefaultendpunct}{\mcitedefaultseppunct}\relax
\EndOfBibitem
\bibitem[Park \latin{et~al.}(2018)Park, Park, Kim, Kim, Kim, and
  Joo]{park2018hybrid}
Park,~H.~J.; Park,~C.-J.; Kim,~J.~Y.; Kim,~M.~S.; Kim,~J.; Joo,~J. Hybrid
  characteristics of MoS2 monolayer with organic semiconducting tetracene and
  application to anti-ambipolar field effect transistor. \emph{ACS applied
  materials \& interfaces} \textbf{2018}, \emph{10}, 32556--32566\relax
\mciteBstWouldAddEndPuncttrue
\mciteSetBstMidEndSepPunct{\mcitedefaultmidpunct}
{\mcitedefaultendpunct}{\mcitedefaultseppunct}\relax
\EndOfBibitem
\bibitem[Amsterdam \latin{et~al.}(2020)Amsterdam, LaMountain, Stanev, Sangwan,
  L{\'o}pez-Arteaga, Padgaonkar, Watanabe, Taniguchi, Weiss, Marks, Hersam, and
  Stern]{amsterdam2020tailoring}
Amsterdam,~S.~H.; LaMountain,~T.; Stanev,~T.~K.; Sangwan,~V.~K.;
  L{\'o}pez-Arteaga,~R.; Padgaonkar,~S.; Watanabe,~K.; Taniguchi,~T.;
  Weiss,~E.~A.; Marks,~T.~J.; Hersam,~M.~C.; Stern,~N.~P. Tailoring the Optical
  Response of Pentacene Thin Films via Templated Growth on Hexagonal Boron
  Nitride. \emph{The Journal of Physical Chemistry Letters} \textbf{2020},
  \emph{12}, 26--31\relax
\mciteBstWouldAddEndPuncttrue
\mciteSetBstMidEndSepPunct{\mcitedefaultmidpunct}
{\mcitedefaultendpunct}{\mcitedefaultseppunct}\relax
\EndOfBibitem
\bibitem[Rotter \latin{et~al.}(2016)Rotter, Lechner, Morherr, Chisnall, Ward,
  Jardine, Ellis, Allison, Eckhardt, and Witte]{rotter2016coupling}
Rotter,~P.; Lechner,~B.~A.; Morherr,~A.; Chisnall,~D.~M.; Ward,~D.~J.;
  Jardine,~A.~P.; Ellis,~J.; Allison,~W.; Eckhardt,~B.; Witte,~G. Coupling
  between diffusion and orientation of pentacene molecules on an organic
  surface. \emph{Nature materials} \textbf{2016}, \emph{15}, 397--400\relax
\mciteBstWouldAddEndPuncttrue
\mciteSetBstMidEndSepPunct{\mcitedefaultmidpunct}
{\mcitedefaultendpunct}{\mcitedefaultseppunct}\relax
\EndOfBibitem
\bibitem[Witte and W{\"o}ll(2004)Witte, and W{\"o}ll]{witte2004growth}
Witte,~G.; W{\"o}ll,~C. Growth of aromatic molecules on solid substrates for
  applications in organic electronics. \emph{Journal of Materials Research}
  \textbf{2004}, \emph{19}, 1889--1916\relax
\mciteBstWouldAddEndPuncttrue
\mciteSetBstMidEndSepPunct{\mcitedefaultmidpunct}
{\mcitedefaultendpunct}{\mcitedefaultseppunct}\relax
\EndOfBibitem
\bibitem[El~Helou \latin{et~al.}(2010)El~Helou, Medenbach, and
  Witte]{el2010rubrene}
El~Helou,~M.; Medenbach,~O.; Witte,~G. Rubrene microcrystals: A route to
  investigate surface morphology and bulk anisotropies of organic
  semiconductors. \emph{Crystal growth \& design} \textbf{2010}, \emph{10},
  3496--3501\relax
\mciteBstWouldAddEndPuncttrue
\mciteSetBstMidEndSepPunct{\mcitedefaultmidpunct}
{\mcitedefaultendpunct}{\mcitedefaultseppunct}\relax
\EndOfBibitem
\bibitem[K{\"a}fer and Witte(2007)K{\"a}fer, and Witte]{kafer2007evolution}
K{\"a}fer,~D.; Witte,~G. Evolution of pentacene films on Ag (1 1 1): Growth
  beyond the first monolayer. \emph{Chemical physics letters} \textbf{2007},
  \emph{442}, 376--383\relax
\mciteBstWouldAddEndPuncttrue
\mciteSetBstMidEndSepPunct{\mcitedefaultmidpunct}
{\mcitedefaultendpunct}{\mcitedefaultseppunct}\relax
\EndOfBibitem
\bibitem[Niederhausen \latin{et~al.}(2020)Niederhausen, MacQueen, Lips,
  Aldahhak, Schmidt, and Gerstmann]{niederhausen2020tetracene}
Niederhausen,~J.; MacQueen,~R.~W.; Lips,~K.; Aldahhak,~H.; Schmidt,~W.~G.;
  Gerstmann,~U. Tetracene ultrathin film growth on hydrogen-passivated silicon.
  \emph{Langmuir} \textbf{2020}, \emph{36}, 9099--9113\relax
\mciteBstWouldAddEndPuncttrue
\mciteSetBstMidEndSepPunct{\mcitedefaultmidpunct}
{\mcitedefaultendpunct}{\mcitedefaultseppunct}\relax
\EndOfBibitem
\bibitem[Breuer \latin{et~al.}(2016)Breuer, Ma{\ss}meyer, M{\"a}nz, Zoerb,
  Harbrecht, and Witte]{breuer2016structure}
Breuer,~T.; Ma{\ss}meyer,~T.; M{\"a}nz,~A.; Zoerb,~S.; Harbrecht,~B.; Witte,~G.
  Structure of van der Waals bound hybrids of organic semiconductors and
  transition metal dichalcogenides: the case of acene films on MoS2.
  \emph{physica status solidi (RRL)--Rapid Research Letters} \textbf{2016},
  \emph{10}, 905--910\relax
\mciteBstWouldAddEndPuncttrue
\mciteSetBstMidEndSepPunct{\mcitedefaultmidpunct}
{\mcitedefaultendpunct}{\mcitedefaultseppunct}\relax
\EndOfBibitem
\bibitem[Gavrila \latin{et~al.}(2004)Gavrila, Mendez, Kampen, Zahn, Vyalikh,
  and Braun]{gavrila2004energy}
Gavrila,~G.; Mendez,~H.; Kampen,~T.; Zahn,~D.; Vyalikh,~D.; Braun,~W. Energy
  band dispersion in well ordered N, N'-dimethyl-3, 4, 9,
  10-perylenetetracarboxylic diimide films. \emph{Applied physics letters}
  \textbf{2004}, \emph{85}, 4657--4659\relax
\mciteBstWouldAddEndPuncttrue
\mciteSetBstMidEndSepPunct{\mcitedefaultmidpunct}
{\mcitedefaultendpunct}{\mcitedefaultseppunct}\relax
\EndOfBibitem
\bibitem[Fedorov(2017)]{fedorov2017first}
Fedorov,~I.~A. First-principles study of band structures of anthracene and
  tetracene under pressure. \emph{Materials Chemistry and Physics}
  \textbf{2017}, \emph{199}, 173--178\relax
\mciteBstWouldAddEndPuncttrue
\mciteSetBstMidEndSepPunct{\mcitedefaultmidpunct}
{\mcitedefaultendpunct}{\mcitedefaultseppunct}\relax
\EndOfBibitem
\bibitem[Cocchi \latin{et~al.}(2018)Cocchi, Breuer, Witte, and
  Draxl]{cocchi2018polarized}
Cocchi,~C.; Breuer,~T.; Witte,~G.; Draxl,~C. Polarized absorbance and Davydov
  splitting in bulk and thin-film pentacene polymorphs. \emph{Physical
  Chemistry Chemical Physics} \textbf{2018}, \emph{20}, 29724--29736\relax
\mciteBstWouldAddEndPuncttrue
\mciteSetBstMidEndSepPunct{\mcitedefaultmidpunct}
{\mcitedefaultendpunct}{\mcitedefaultseppunct}\relax
\EndOfBibitem
\bibitem[Doi \latin{et~al.}(2005)Doi, Yoshida, Nakano, Tachibana, Tanabe,
  Kojima, and Okazaki]{doi2005ab}
Doi,~K.; Yoshida,~K.; Nakano,~H.; Tachibana,~A.; Tanabe,~T.; Kojima,~Y.;
  Okazaki,~K. Ab initio calculation of electron effective masses in solid
  pentacene. \emph{Journal of applied physics} \textbf{2005}, \emph{98},
  113709\relax
\mciteBstWouldAddEndPuncttrue
\mciteSetBstMidEndSepPunct{\mcitedefaultmidpunct}
{\mcitedefaultendpunct}{\mcitedefaultseppunct}\relax
\EndOfBibitem
\bibitem[Bergh\"auser and Malic(2014)Bergh\"auser, and Malic]{gunnar_prb}
Bergh\"auser,~G.; Malic,~E. Analytical approach to excitonic properties of
  MoS${}_{2}$. \emph{Phys. Rev. B} \textbf{2014}, \emph{89}, 125309\relax
\mciteBstWouldAddEndPuncttrue
\mciteSetBstMidEndSepPunct{\mcitedefaultmidpunct}
{\mcitedefaultendpunct}{\mcitedefaultseppunct}\relax
\EndOfBibitem
\bibitem[Brem \latin{et~al.}(2018)Brem, Selig, Berghaeuser, and
  Malic]{brem2018exciton}
Brem,~S.; Selig,~M.; Berghaeuser,~G.; Malic,~E. Exciton Relaxation Cascade in
  two-dimensional Transition Metal Dichalcogenides. \emph{Scientific reports}
  \textbf{2018}, \emph{8}, 8238\relax
\mciteBstWouldAddEndPuncttrue
\mciteSetBstMidEndSepPunct{\mcitedefaultmidpunct}
{\mcitedefaultendpunct}{\mcitedefaultseppunct}\relax
\EndOfBibitem
\bibitem[Kormanyos \latin{et~al.}(2015)Kormanyos, Burkard, Gmitra, Fabian,
  Zolyomi, Drummond, and Falko]{andor}
Kormanyos,~A.; Burkard,~G.; Gmitra,~M.; Fabian,~J.; Zolyomi,~V.;
  Drummond,~N.~D.; Falko,~V. k · p theory for two-dimensional transition metal
  dichalcogenide semiconductors. \emph{2D Materials} \textbf{2015}, \emph{2},
  022001\relax
\mciteBstWouldAddEndPuncttrue
\mciteSetBstMidEndSepPunct{\mcitedefaultmidpunct}
{\mcitedefaultendpunct}{\mcitedefaultseppunct}\relax
\EndOfBibitem
\bibitem[Hagel \latin{et~al.}(2021)Hagel, Brem, Linderälv, Erhart, and
  Malic]{hagel2021}
Hagel,~J.; Brem,~S.; Linderälv,~C.; Erhart,~P.; Malic,~E. Exciton landscape in
  van der Waals heterostructures. \emph{arXiv preprint arXiv: 2109.10746}
  \textbf{2021}, \relax
\mciteBstWouldAddEndPunctfalse
\mciteSetBstMidEndSepPunct{\mcitedefaultmidpunct}
{}{\mcitedefaultseppunct}\relax
\EndOfBibitem
\bibitem[Chernikov \latin{et~al.}(2014)Chernikov, Berkelbach, Hill, Rigosi, Li,
  Aslan, Reichman, Hybertsen, and Heinz]{Chernikov2014}
Chernikov,~A.; Berkelbach,~T.~C.; Hill,~H.~M.; Rigosi,~A.; Li,~Y.;
  Aslan,~O.~B.; Reichman,~D.~R.; Hybertsen,~M.~S.; Heinz,~T.~F. Exciton Binding
  Energy and Nonhydrogenic Rydberg Series in Monolayer ${\mathrm{WS}}_{2}$.
  \emph{Phys. Rev. Lett.} \textbf{2014}, \emph{113}, 076802\relax
\mciteBstWouldAddEndPuncttrue
\mciteSetBstMidEndSepPunct{\mcitedefaultmidpunct}
{\mcitedefaultendpunct}{\mcitedefaultseppunct}\relax
\EndOfBibitem
\bibitem[Arora \latin{et~al.}(2015)Arora, Koperski, Nogajewski, Marcus,
  Faugeras, and Potemski]{arora2015excitonic}
Arora,~A.; Koperski,~M.; Nogajewski,~K.; Marcus,~J.; Faugeras,~C.; Potemski,~M.
  Excitonic resonances in thin films of WSe 2: from monolayer to bulk material.
  \emph{Nanoscale} \textbf{2015}, \emph{7}, 10421--10429\relax
\mciteBstWouldAddEndPuncttrue
\mciteSetBstMidEndSepPunct{\mcitedefaultmidpunct}
{\mcitedefaultendpunct}{\mcitedefaultseppunct}\relax
\EndOfBibitem
\bibitem[Mueller and Malic(2018)Mueller, and Malic]{erminnpj}
Mueller,~T.; Malic,~E. Exciton physics and device application of
  two-dimensional transition metal dichalcogenide semiconductors. \emph{npj 2D
  Materials and Applications} \textbf{2018}, \emph{2}, 1--12\relax
\mciteBstWouldAddEndPuncttrue
\mciteSetBstMidEndSepPunct{\mcitedefaultmidpunct}
{\mcitedefaultendpunct}{\mcitedefaultseppunct}\relax
\EndOfBibitem
\bibitem[Feldtmann \latin{et~al.}(2009)Feldtmann, Kira, and
  Koch]{feldtmann2009phonon}
Feldtmann,~T.; Kira,~M.; Koch,~S.~W. Phonon sidebands in semiconductor
  luminescence. \emph{physica status solidi (b)} \textbf{2009}, \emph{246},
  332--336\relax
\mciteBstWouldAddEndPuncttrue
\mciteSetBstMidEndSepPunct{\mcitedefaultmidpunct}
{\mcitedefaultendpunct}{\mcitedefaultseppunct}\relax
\EndOfBibitem
\bibitem[Chernikov \latin{et~al.}(2012)Chernikov, Bornwasser, Koch, Chatterjee,
  B{\"o}ttge, Feldtmann, Kira, Koch, Wassner, Lautenschl{\"a}ger,
  \latin{et~al.} others]{chernikov2012phonon}
Chernikov,~A.; Bornwasser,~V.; Koch,~M.; Chatterjee,~S.; B{\"o}ttge,~C.;
  Feldtmann,~T.; Kira,~M.; Koch,~S.; Wassner,~T.; Lautenschl{\"a}ger,~S.,
  \latin{et~al.}  Phonon-assisted luminescence of polar semiconductors:
  Fr{\"o}hlich coupling versus deformation-potential scattering. \emph{Physical
  Review B} \textbf{2012}, \emph{85}, 035201\relax
\mciteBstWouldAddEndPuncttrue
\mciteSetBstMidEndSepPunct{\mcitedefaultmidpunct}
{\mcitedefaultendpunct}{\mcitedefaultseppunct}\relax
\EndOfBibitem
\bibitem[Axt \latin{et~al.}(1996)Axt, Victor, and Stahl]{axt1996influence}
Axt,~V.; Victor,~K.; Stahl,~A. Influence of a phonon bath on the hierarchy of
  electronic densities in an optically excited semiconductor. \emph{Physical
  Review B} \textbf{1996}, \emph{53}, 7244\relax
\mciteBstWouldAddEndPuncttrue
\mciteSetBstMidEndSepPunct{\mcitedefaultmidpunct}
{\mcitedefaultendpunct}{\mcitedefaultseppunct}\relax
\EndOfBibitem
\bibitem[Selig \latin{et~al.}(2016)Selig, Bergh{\"a}user, Raja, Nagler,
  Sch{\"u}ller, Heinz, Korn, Chernikov, Malic, and Knorr]{selig2016excitonic}
Selig,~M.; Bergh{\"a}user,~G.; Raja,~A.; Nagler,~P.; Sch{\"u}ller,~C.;
  Heinz,~T.~F.; Korn,~T.; Chernikov,~A.; Malic,~E.; Knorr,~A. Excitonic
  linewidth and coherence lifetime in monolayer transition metal
  dichalcogenides. \emph{Nature communications} \textbf{2016}, \emph{7},
  13279\relax
\mciteBstWouldAddEndPuncttrue
\mciteSetBstMidEndSepPunct{\mcitedefaultmidpunct}
{\mcitedefaultendpunct}{\mcitedefaultseppunct}\relax
\EndOfBibitem
\bibitem[Rosati \latin{et~al.}(2020)Rosati, Wagner, Brem, Perea-Causín,
  Wietek, Zipfel, Ziegler, Selig, Taniguchi, Watanabe, Knorr, Chernikov, and
  Malic]{doi:10.1021/acsphotonics.0c00866}
Rosati,~R.; Wagner,~K.; Brem,~S.; Perea-Causín,~R.; Wietek,~E.; Zipfel,~J.;
  Ziegler,~J.~D.; Selig,~M.; Taniguchi,~T.; Watanabe,~K.; Knorr,~A.;
  Chernikov,~A.; Malic,~E. Temporal Evolution of Low-Temperature Phonon
  Sidebands in Transition Metal Dichalcogenides. \emph{ACS Photonics}
  \textbf{2020}, \emph{7}, 2756--2764\relax
\mciteBstWouldAddEndPuncttrue
\mciteSetBstMidEndSepPunct{\mcitedefaultmidpunct}
{\mcitedefaultendpunct}{\mcitedefaultseppunct}\relax
\EndOfBibitem
\bibitem[Varshni(1967)]{varshni1967temperature}
Varshni,~Y.~P. Temperature dependence of the energy gap in semiconductors.
  \emph{physica} \textbf{1967}, \emph{34}, 149--154\relax
\mciteBstWouldAddEndPuncttrue
\mciteSetBstMidEndSepPunct{\mcitedefaultmidpunct}
{\mcitedefaultendpunct}{\mcitedefaultseppunct}\relax
\EndOfBibitem
\end{mcitethebibliography}
\providecommand{\latin}[1]{#1}
\makeatletter
\providecommand{\doi}
  {\begingroup\let\do\@makeother\dospecials
  \catcode`\{=1 \catcode`\}=2 \doi@aux}
\providecommand{\doi@aux}[1]{\endgroup\texttt{#1}}
\makeatother
\providecommand*\mcitethebibliography{\thebibliography}
\csname @ifundefined\endcsname{endmcitethebibliography}
  {\let\endmcitethebibliography\endthebibliography}{}

\end{document}


\title{Supplementary Information: Interlayer exciton landscape in WS$_2$/tetracene heterostructures}

\author{Joshua J. P. Thompson}
\affiliation{ Department of Physics, Philipps-Universität Marburg,
35037 Marburg, Germany}
\author{Victoria Lumsargis}
\affiliation{Department of Chemistry, Purdue University, West Lafayette, Indiana, 47907, United States }
\author{Maja Feierabend}
\affiliation{ Department of 
Physics, Chalmers University of Technology,  412 96 Gothenburg, Sweden}
\author{Quichen Zhao}
\affiliation{Department of Chemistry, Purdue University, West Lafayette, Indiana, 47907, United States }
\affiliation{State Key Laboratory of Superhard Materials, Jilin University, Changchun, Jilin, 130012, China}
\author{Kang Wang}
\affiliation{ Davidson School of Chemical Engineering,  Purdue University, West Lafayette, Indiana,  47907, United States}
\author{Letian Dou}
\affiliation{ Davidson School of Chemical Engineering,  Purdue University, West Lafayette, Indiana,  47907, United States}
\author{Libai Huang}
\affiliation{Department of Chemistry, Purdue University, West Lafayette, Indiana, 47907, United States }
\author{Ermin Malic}
\affiliation{ Department of Physics, Philipps-Universität Marburg,
35037 Marburg, Germany}
\affiliation{ Department of 
Physics, Chalmers University of Technology,  412 96 Gothenburg, Sweden}

\maketitle

\section{Preparation of tungsten disulfide/tetracene heterostructure
}
\subsection{Tungsten disulfide preparation}
The silicon dioxide wafer (oxide thickness of 90 nm) was cleaned in a sonication bath. The wafers were first sonicated in acetone, then isopropyl alcohol and finally deionized water for five minutes each. Bulk tungsten disulfide (WS$_2$) crystals (purchased from Graphene Supermarket) were then exfoliated onto the cleaned wafer using Scotch Magic Tape. Before lifting off the tape from the wafer, the bulk crystal taped to the silicon wafer was annealed on a hot plate at 100$^\circ$C for two minutes \cite{huang2015reliable}. The monolayer of WS$_2$ was found on the substrate using an Olympus Fluorescence Microscope. Optical images of the monolayer are shown in Figure S1a. A mercury lamp was used to excite the monolayer which is shown in red in Figure S1a(ii). The room temperature PL spectra (447nm excitation) of the monolayer, heterostructure and pure tetracene used in this study is shown in Figure S1b. The single, narrow KK exciton peak at 2.0 eV is indicative that the WS$_2$ flake is a monolayer. The inset in Figure S1b is an enlargement of the spectra between 1.4-2.0 eV.

\begin{figure}[!h]
    \centering
    \includegraphics[width = 0.9\linewidth]{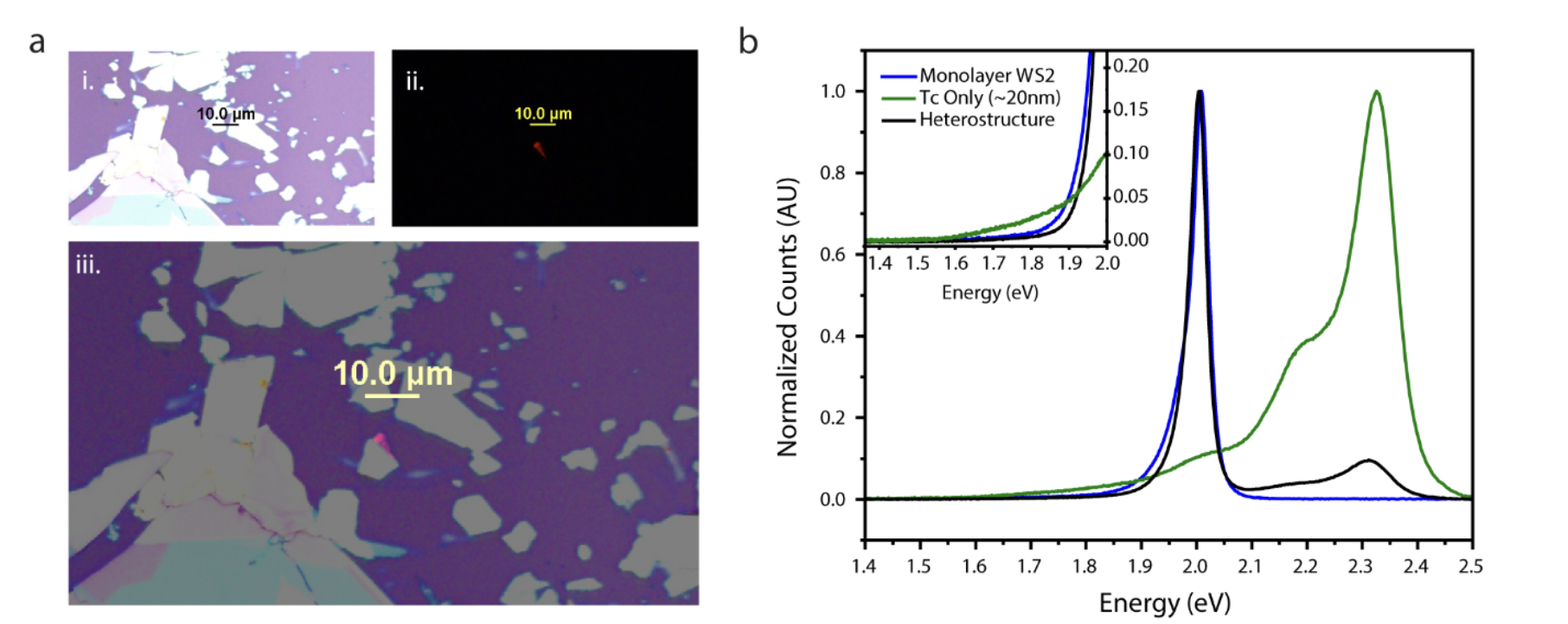}
    \caption{Characterization of WS$_2$, Tc and the heterostructure. (a) Optical Images of (i) the exfoliated monolayer of WS$_2$ (ii) the monolayer being excited with a mercury lamp (iii) the overlay of images (i) and (ii). (b) the PL spectra of monolayer WS$_2$, pure Tc and the heterostructure when excited with 447 nm light. The inset in (b) is an enlargement of the main spectra between 1.4-2.0 eV. }
    \label{S1}
\end{figure}

\subsection{Tetracene Evaporation}

Tetracene (98$\%$ ACROS Organics) was then thermally evaporated onto the sample using a LC Technology thermal evaporator with an INFICON (SQC-310) deposition controller. The tetracene was deposited at a rate of 0.5\AA/s under a vacuum of $<$ 2 x 10$^{-6}$ mbar. The tetracene (Tc) height was measured using a Veeco Dimension 3100 Atomic Force Microscope (AFM). The AFM image in Figure S2 shows two profiles which were used to determine a tetracene thickness of $\sim$ 20 nm.

\begin{figure}[!h]
    \centering
    \includegraphics[width = 0.9\linewidth]{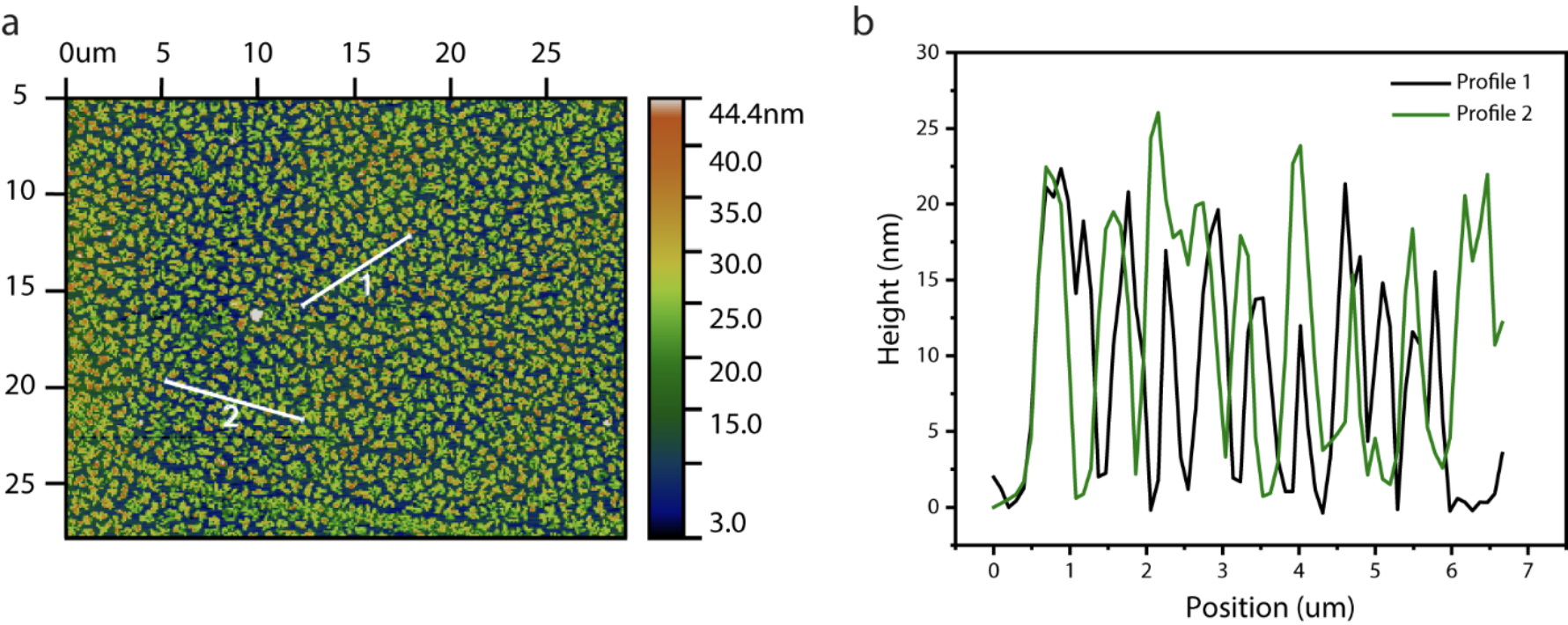}
    \caption{AFM of Tc. (a) AFM image of pure tetracene showing the location of the two profiles plotted in (b).}
    \label{S2}
\end{figure}

\section{Low Temperature Photoluminescence Measuerments}

To measure the low temperature spectra of the heterostructure, the sample was cooled using a Montana Instrument Cryostation s50 (temperature stability $<$ 10 mK). The steady state PL spectra were collected using a home-built confocal micro-PL set up as described in previous publications \cite{zhu2018highly}. The sample was excited using a PHAROS Light Conversion Ltd. Laser (repetition rate of 750 kHz) whose output was directed into an optical parametric amplifier (OPA, TOPAS-Twins, Light Conversion Ltd) to create 580 nm (2.1377 eV) light. This wavelength was chosen to selectively excite the WS$_2$ and minimise tetracene emission, as shown in the inset of Figure S1b, when using 447 nm excitation, Tc emission overpowers that of the charge transfer (CT) exciton in the heterostructure, as no CT peak is present at 1.7 eV. The 580 nm light (fluence of 112 $\mu$Jcm$^{-2}$) was focused using a 40x numerical aperture (NA) 0.6 objective. The sample emission was collected using the same objective and later dispersed and detected using an Andor Technology monochromator and thermoelectric-cooled charge-coupled device. The lifetime measurements included in Figure S3 were collected using a single photon avalanche diode (Pico-Quant, PDM series) and a single photon counting module (Pico-Quant) with a time resolution of $\sim$ 100 ps.

\begin{figure}[!h]
    \centering
    \includegraphics[width = 0.5\linewidth]{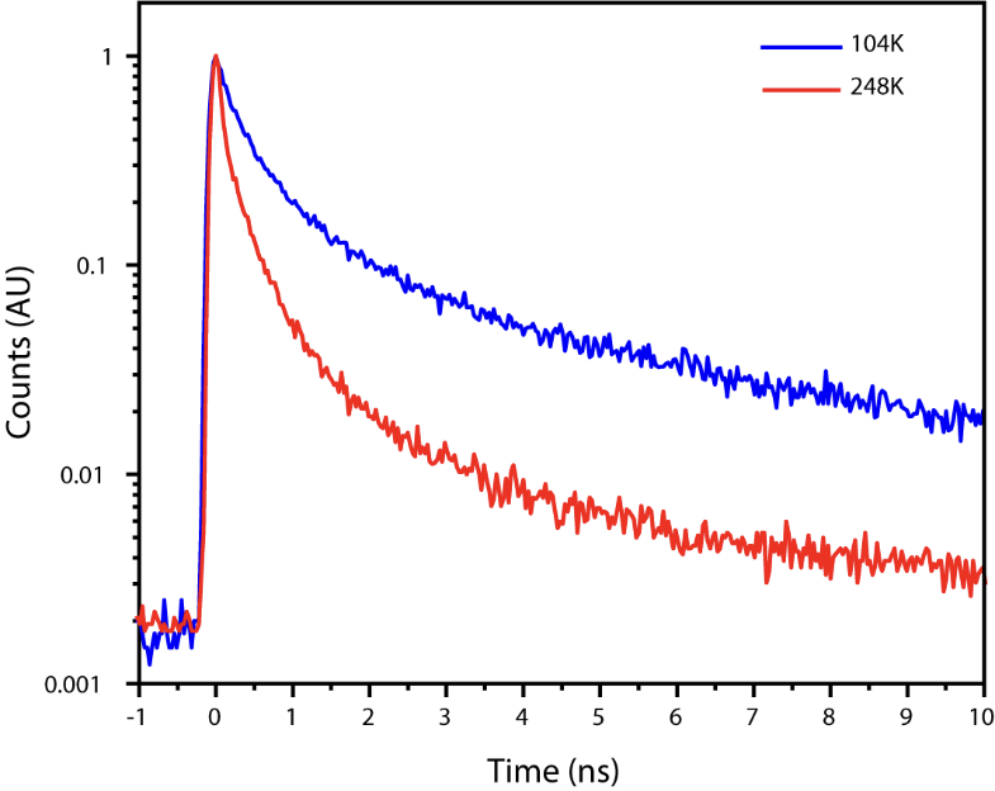}
    \caption{ Lifetime dynamics of the heterostructure excited with 580 nm light at 104 K and 248 K}
    \label{S3}
\end{figure}
In addition to the PL measurements presented in the main text, we performed additional measurements to verify that the observed signatures persist in multiple experiments. For a more direct comparison between the experimental trials, we subtract the background emission to isolate the pertinent features in the PL of the heterostructure. In Figure S4 (a) we show the results presented in the main text only with background subtraction applied. We can compare this to an additional trial, shown in Figure S4b and we observe good agreement both in the peak intensities and positions.
\begin{figure}[!h]
    \centering
    \includegraphics[width = 0.85\linewidth]{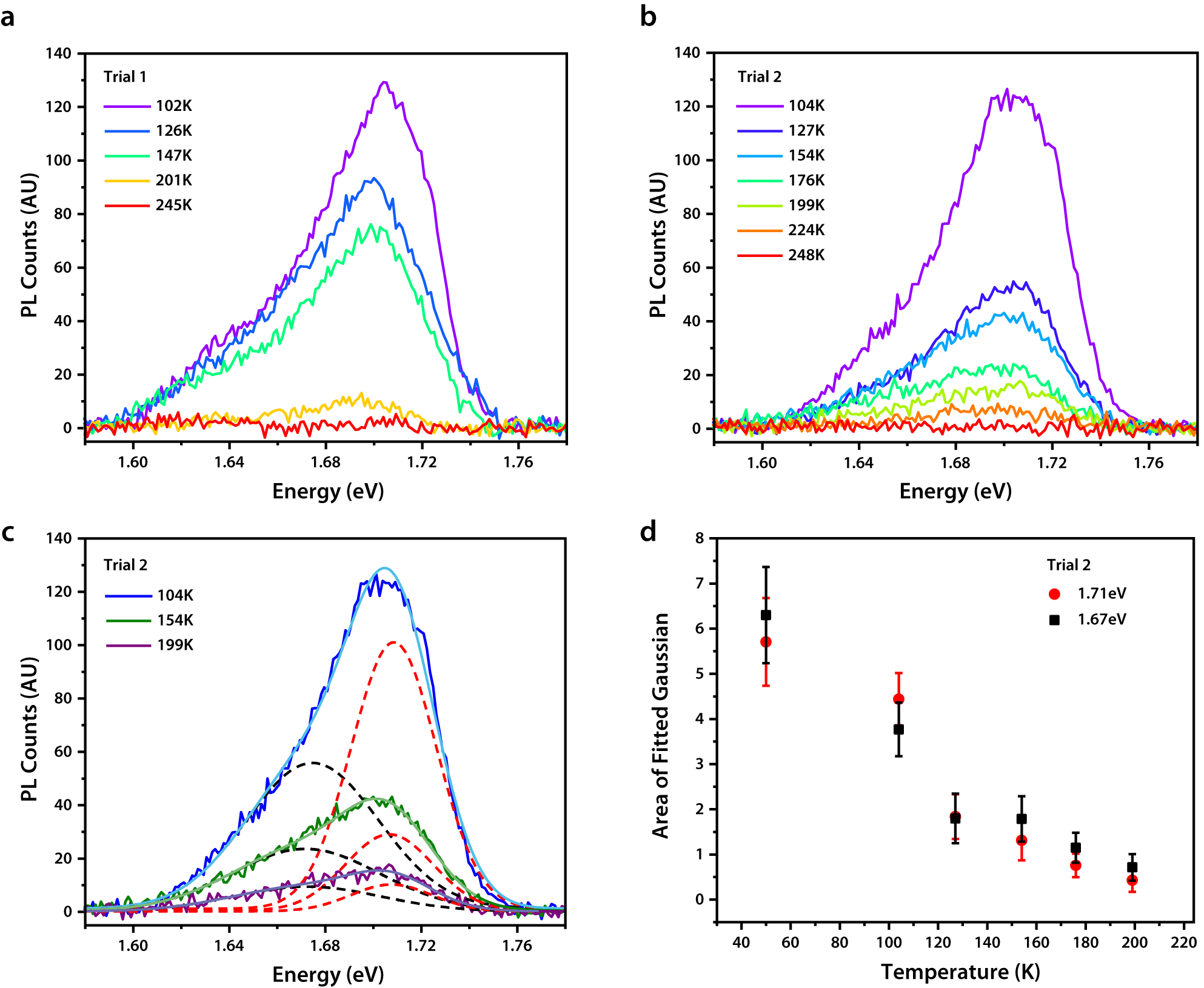}
    \caption{ PL spectra following background subtraction for (a) Trial 1 (as shown in main text) and (b) Trial 2, at different temperatures. (c) Double-Gaussian fit of Trial 2 for different temperatures. (d) Temperature dependence of Gaussian fit peaks at 1.67 eV (black) and 1.7 eV (red) presented in (c). }
    \label{S4}
\end{figure}

In Figure S4c, we present double-Gaussian fits of the experimental data shown in (b), supporting the idea that the observed PL spectra can be attributed to an interlayer exciton hK’ and a phonon-sideband with Gaussian peaks at 1.67 eV to 1.7 eV. The peak at 1.7 eV we assign to the hK’ exciton while the broader peak at 1.67 eV, we propose is formed from multiple phonon-sidebands at different energy, leading to a broad curve.

In our model, the change in peak intensity with temperature could be attributed to either a change in the relative thermal occupation of these excitons, $N_\mu$, or the broadening of the peaks by increased phonon-scattering. By integrating the area of the peaks (as in Figure S4c), we can attribute the oscillator strength transfer described in the main text to a change in the thermal occupation (as predicted by our model). This is presented in Figure S4d.

\begin{figure}[!h]
    \centering
    \includegraphics[width = 0.85\linewidth]{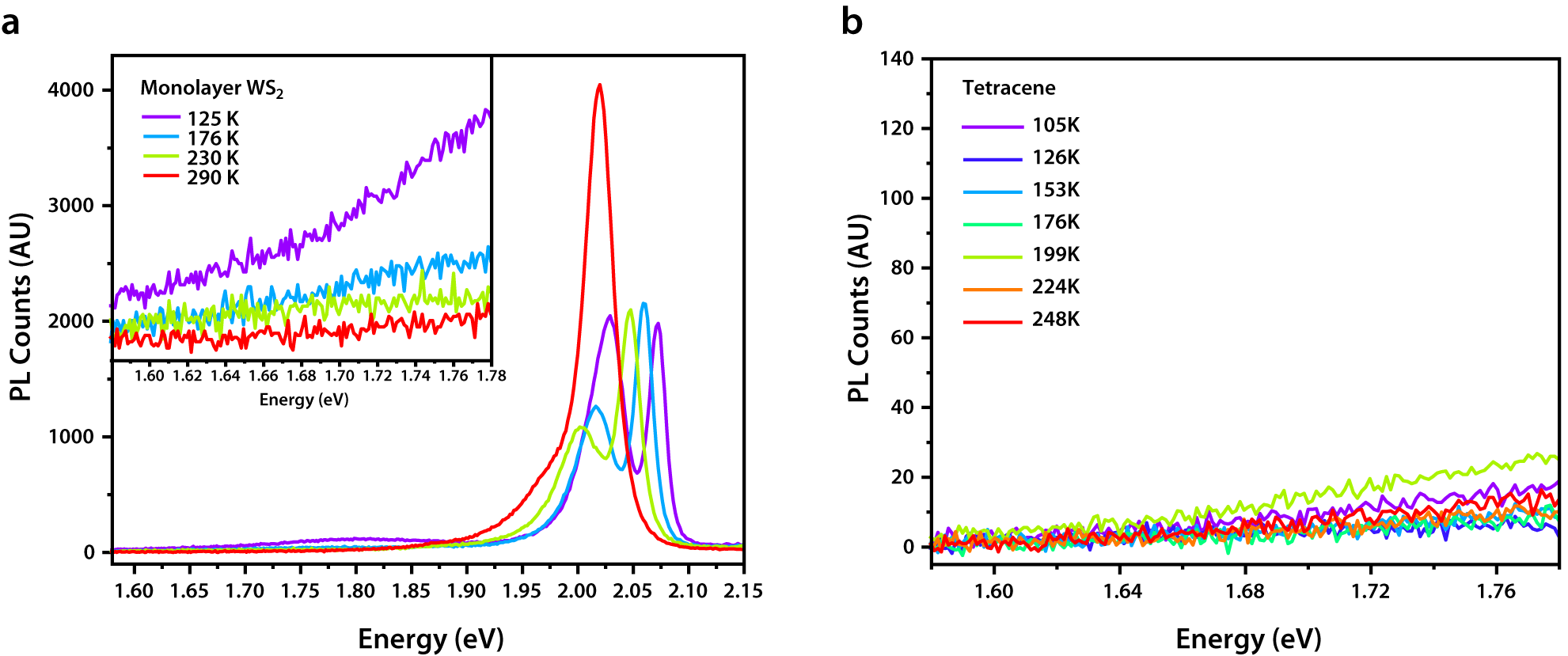}
    \caption{(a) Temperature dependent PL spectra of monolayer WS$_2$ excited with 447 nm light. The inset in (a) is an enlargement of the spectra between 1.58-1.78 eV. (b) Temperature dependent PL spectra of tetracene excited with 580 nm light.}
    \label{S5}
\end{figure}

Furthermore, the temperature dependent PL was collected for a monolayer of WS$_2$ and pure Tc. In Figure S5a, we display the temperature dependent PL of monolayer WS$_2$ excited with 447 nm light. The inset shows the absence of the interlayer exciton $hK’$ and phonon-sideband between 1.58-1.78 eV. Likewise, Figure S5b shows the temperature dependent PL spectrum of pure Tc excited with 580nm light and also demonstrates the absence of the peaks of interest. This confirms that the signals present in Figure S4a-c are unique to the heterostructure. 

\section{Theoretical Model}
\subsection{Wannier Equation}
In order to determine the excitonic binding energies and wavefunctions we employ the Wannier equation \cite{Kira2006}
\begin{align}
\left(\dfrac{\hbar^2 \vect{q}^2}{2 m^r_{ij}} +E^{\text{Gap}}_{ij}\right)\varphi^{ij, \mu}_{\vect{q}} + \sum_{\vect{k}}V^{ij}(\vect{k}+\vect{q})\varphi^{ij, \mu}_{\vect{k}} = E^b_{ij, \mu} \varphi^{ij, \mu}_{\vect{q}}
\end{align}
where $\vect{q}$ and $\vect{k}$ are  momenta and $E^{\text{Gap}}_{ij}$ is the band gap energy between an electron in layer/valley $i$ and hole in layer/valley $j$. In this equation we also have the reduced mass $m^r_{ij} = m^e_i m^h_j/ (m^e_i+m^h_j)$  and the screened Coulomb interaction $V^{ij}(\vect{k}+\vect{q})$ between the electron and hole which varies for inter/intra- layer interaction \cite{ovesen2019interlayer}. The excitonic wavefunction is $\varphi^{ij, \mu}_{\vect{q}}$ where $\mu$ signifies the excitonic energy level, resembling those of a hydrogen atom (1s, 2s, 2p etc). In this work we only consider the lowest-lying, optically active 1s exciton states, hence we drop the index $\mu$ and use it to represent the combined indices $\mu = (i,j)$ determining whether an exciton is inter/intralayer as well as the valley indices. $E^b_{ij, \mu}$ are the corresponding binding energies.

\subsection{Elliot Formula}
In this section we aim to outline the origin of the excitonic Elliot formula in a TMD/Tc heterostructure within our density matrix formalism. The PL intensity can be described using the time evolution of the photon number, $I_\text{PL}(\omega_q)=\omega_q \frac{\partial}{\partial t} n_q\propto S_{\bf{Q}}^{\mu}$ with  the number of emitted photons $ n_q=\langle c^{\dagger}_qc_q\rangle$ and the photon-assisted polarization $S_{\bf{Q}}^{\mu} = \sum_{q} \varphi_{\bf q}^{\mu *} S_{\bf{qQ}}^{vc }$ with  $S^{v c}_{\bf{k_1}\bf{k_2}}(t)=\langle c_q^\dagger a_{\bf k_1}^{\dagger v} a_{\bf k_2}^{c} \rangle$ 
 \cite{brem2019phonon}. 
  This microscopic quantity describes optically induced transitions from an initial state $(v,\boldsymbol{ k_1})$ to a final state $(c,\boldsymbol{k_2}) $ under annihilation (creation) of a photon $c_q^{(\dagger)}$. 
  Here, $\bf k_i$ is the electronic momentum and $\lambda=v,c$ the band index denoting the valence or conduction band, respectively. 

Next, we exploit the Heisenberg equation of motion $i\hbar \dot S_{\bf Q}^\mu (t)=[H,S_{\bf Q}^\mu (t)]$ \cite{Kochbuch} to obtain the temporal evolution of the photon-assisted polarization. The Hamilton operator
$H=H_{\text{0}}+H_{\text{x-phot}}+H_{\text{x-phon}}+H_{\text{x-tunn}}$
includes the interaction-free part for excitons, photons and phonons $H_{\text{0}}$, the exciton-photon interaction $H_{\text{x-phot}}$  including the optical matrix element $ M^{i\sigma }_{Q}$, and  the exciton-phonon coupling $H_{\text{x-phon}}$. 
The crucial contribution to the Hamilton operator in this work is the tunneling part $H_{\text{tunn}}=\sum_{QQ',ij}
    T^{ij}_{QQ'} X^{\dagger i}_{Q} X^{j}_{Q-Q'}$ describing the charge transfer between the layers. The appearing tunneling matrix element $T^{ij}_{QQ'} = \sum_q \varphi_{q+\alpha Q}^{i} t^{ij}_{QQ'} \varphi_{q-\beta Q'}^{j}$ is determined by the overlap of Bloch waves $\Psi_i$ with a tunneling potential $V_{\text{tun}} $, i.e. 
$t^{ij}=\langle\Psi_{i} | V_{\text{tun}} |  \Psi_{j} \rangle$. We  separate the potential into an in-plane disorder part and out-of-plane part, which is only non-zero between the layers \cite{ovesen2019interlayer}. Then, we can write $t^{ij} = V_{in}(|k_j-k_i|) V_{out} \langle {u^{i}|u^{j} \rangle}_{uc}$ with $u^{i}$ the lattice periodic parts of the Bloch waves integrated over a unit cell. This integral determines the tunneling strength and we assume a typical value of 0.01 known from   TMD/TMD heterostructures \cite{ovesen2019interlayer}. The remaining in-plane disorder potential can be estimated via a disorder potential with correlation length in the order of the exciton Bohr radius.

In a last step, we project into a new basis where the tunneling part of the Hamiltonian is included in an effective single-particle Hamiltonian, i.e. \begin{align}
    H_0+H_{\text{x-tun}} = \sum_{Q,i}
    \epsilon^{i}_{Q} X^{\dagger i}_{Q} X^{i}_{Q} +
\sum_{QQ',ij}
    T^{ij}_{QQ'} X^{\dagger i}_{Q} X^{j}_{Q-Q'} = \sum_{\mu Q} \mathcal{E}_{\mu Q}\tilde{X}^{\dagger }_{\mu Q} \tilde{X}^{}_{\mu Q}.
\end{align} 
Note that for the applied transformation $\hat{U}_{\mu}$, we exploited
$\tilde{X}^{\dagger}_{\mu Q}
=\hat{U}_{\mu} X^{ \dagger}_{\mu Q}= \sum_j  U^{j}_{\mu Q } X^{j \dagger}_{Q}$ and $\hat{U}_{\mu}^{\dagger} \hat{U}_{\mu} = \mathcal{I}$, i.e. a uniform transformation. As intra- and interlayer states are off-resonant and we do not include hybridization effects, eigenenergies are not affected by this transformation. Solving Heisenberg's equations of motion in this excitonic tunnelling basis,  we finally arrive at the Elliot formula for the phonon assisted polarisation $ I^{b}_{\text{PL}}$ for intra- and interlayer excitons  \cite{brem2019phonon}  
\begin{equation}
 \label{PL}
  I^{}_{\text{PL}}(\omega)\propto \sum_{\mu =\text{ KK,hK}} \frac{|M_{ \mu}|^2 \gamma_{\mu} N_{\mu}}{ (E_{\mu}-\omega)^2+(\gamma_{\mu}+\Gamma_{\mu })^2},   
 \end{equation}
with terms described in the main text.
\subsection{Energy Landscape and Screening Paramters}
To calculate the appearing matrix elements in the Hamilton operator, we apply a nearest-neighbor tight-binding approach \cite{ermin_cm,Kochbuch,Kira2006} which includes fixed (not adjustable) input parameters (effective masses, band gap energies, cf. Table \ref{table1})  from DFT calculations for the electronic bandstructure \cite{andor,breuer2016structure}.
\begin{table}[h!]
\centering
\begin{tabularx}{.99\textwidth}{X X X X X X X }
\rowcolor{lightgray}
\multicolumn{7}{c}{energies for state $\mu$} \\
&KK 
&hK
&hK'
&$\text{h}\Lambda$
&Kl
&hl \\
$E^g_\mu$ 
&2.18 
&2.01 
&1.98
&2.03
&3.4
&3.0\\
$E^b_{\mu} $
&196
&243
&272
&328
&272
&440 \\
$E_{\mu} $
&2.00
&1.76
&1.705
&1.71
&3.12
&2.56 \\
\end{tabularx}
\begin{tabularx}{.48\textwidth}{X X X X X X }
\rowcolor{lightgray}
\multicolumn{6}{c}{screening parameters} \\
$\epsilon^{\text{TMD}}$
& $\epsilon^{\text{Tc}}$
& $\epsilon^{\text{sur}}$
& $d^{\text{TMD}}$
& $d^{\text{Tc}}$ 
& $d^{\text{TMD-Tc}} $\\
13.61 
& 3.69 
& 2.45 
& 0.61 
& 1.1 
& 0.3 \\
\end{tabularx}
\begin{tabularx}{.48\textwidth}{X X X X X X }
\rowcolor{lightgray}
\multicolumn{6}{c}{electronic masses} \\
$m^c_{K}$
& $m^c_{\Lambda}$
& $m^c_{K'}$
& $m^v_{K}$
& $m^h$
& $m^l$\\
0.27
&0.64
&0.36
&0.36
&3.0 
&6.7 \\
\end{tabularx}
\caption{
\textbf{Energy Landscape and screening parameters.} Electronic band gap energies $E^g_{\mu}$ (in eV, from \cite{zhu2018highly,andor}), calculated exciton binding energies $E^b_{\mu}$ (meV) and spectral position $E_{\mu}$ (eV) for intra- and interlayer excitons, screening parameters for dielectric environment $\epsilon$ (from \cite{laturia2018dielectric,leenaerts2008adsorption}) and thickness $d$ (in nm, from \cite{laturia2018dielectric,xie2019photoinduced}) and electronic masses (in $m_0$, from \cite{andor,doi2005ab}). All values refer to a WS$_2$/Tc heterostructure on an SiO$_2$ substrate.
}
\label{table1}
\end{table}

\bibliographystyle{apsrev4-1}